\documentclass[dvips]{elsart}
\usepackage{amssymb}

\usepackage[german,english]{babel}
\usepackage{epsfig}
\usepackage{times}
\usepackage{graphicx}
\usepackage{upgreek}
\usepackage{subfigure}
\usepackage[nospace]{cite}
\usepackage{color} % just used for reviewing purposes
\usepackage{soul} % underline with line break, strike out, etc.; 
\newcommand{\comment}[1]{}

\usepackage{lineno}

\begin{document}

\begin{frontmatter}

\title{AMADEUS -- The Acoustic Neutrino Detection Test System of the 
ANTARES Deep-Sea Neutrino Telescope}

%\documentclass{elsart}
%\pagestyle{empty}
%\journal{}
%\begin{document}
%\begin{frontmatter}
%\title{Amadeus_paper.}
\author[IFIC]{J.A. Aguilar},
\author[CPPM]{I. Al Samarai},
\author[Colmar]{A. Albert},
\author[Genova]{M. Anghinolfi},
\author[Erlangen]{G. Anton},
\author[IRFU/SEDI]{S. Anvar},
\author[UPV]{M. Ardid},
\author[NIKHEF]{A.C. Assis Jesus},
\author[NIKHEF]{T.~Astraatmadja\thanksref{tag:1}},
\author[CPPM]{J-J. Aubert},
% ...
\corauth[cor:1]{Corresponding author; Tel.: +49\,9131\,8527147}

\newpage
\nopagebreak[3]
\address[IFIC]{\scriptsize{IFIC - Instituto de F\'isica Corpuscular, Edificios Investigaci\'on de Paterna, CSIC - Universitat de Val\`encia, Apdo. de Correos 22085, 46071 Valencia, Spain}}\vspace*{0.15cm}
\nopagebreak[3]
\vspace*{-0.20\baselineskip}
\nopagebreak[3]
\address[CPPM]{\scriptsize{CPPM - Centre de Physique des Particules de Marseille, CNRS/IN2P3 et Universit\'e de la M\'editerran\'ee, 163 Avenue de Luminy, Case 902, 13288 Marseille Cedex 9, France}}\vspace*{0.15cm}
\nopagebreak[3]
\vspace*{-0.20\baselineskip}
\nopagebreak[3]
\address[Colmar]{\scriptsize{GRPHE - Institut universitaire de technologie de Colmar, 34 rue du Grillenbreit BP 50568, 68008 Colmar, France }}\vspace*{0.15cm}
\nopagebreak[3]
\vspace*{-0.20\baselineskip}
\nopagebreak[3]
\address[Genova]{\scriptsize{INFN - Sezione di Genova, Via Dodecaneso 33, 16146 Genova, Italy}}\vspace*{0.15cm}
\nopagebreak[3]
\vspace*{-0.20\baselineskip}
\nopagebreak[3]
\address[Erlangen]{\scriptsize{Friedrich-Alexander-Universit\"{a}t Erlangen-N\"{u}rnberg, Erlangen Centre for Astroparticle Physics, Erwin-Rommel-Str. 1, 91058 Erlangen, Germany}}\vspace*{0.15cm}
\nopagebreak[3]
\vspace*{-0.20\baselineskip}
\nopagebreak[3]
\address[IRFU/SEDI]{\scriptsize{Direction des Sciences de la Mati\`ere - Institut de recherche sur les lois fondamentales de l'Univers - Service d'Electronique des D\'etecteurs et d'Informatique, CEA Saclay, 91191 Gif-sur-Yvette Cedex, France}}\vspace*{0.15cm}
\nopagebreak[3]
\vspace*{-0.20\baselineskip}
\nopagebreak[3]
\address[UPV]{\scriptsize{Institut d'Investigaci\'o per a la Gesti\'o Integrada de Zones Costaneres (IGIC) - Universitat Polit\`ecnica de Val\`encia. C/ Paranimf 1. , 46730 Gandia, Spain.}}\vspace*{0.15cm}
\nopagebreak[3]
\vspace*{-0.20\baselineskip}
\nopagebreak[3]
\address[NIKHEF]{\scriptsize{FOM Instituut voor Subatomaire Fysica Nikhef, Science Park 105, 1098 XG Amsterdam, The Netherlands}}\vspace*{0.15cm}
\nopagebreak[3]
\vspace*{-0.20\baselineskip}
\nopagebreak[3]
\address[Bari]{\scriptsize{INFN - Sezione di Bari, Via E. Orabona 4, 70126 Bari, Italy}}\vspace*{0.15cm}
\nopagebreak[3]
\vspace*{-0.20\baselineskip}
\nopagebreak[3]
\address[APC]{\scriptsize{APC - Laboratoire AstroParticule et Cosmologie, UMR 7164 (CNRS, Universit\'e Paris 7 Diderot, CEA, Observatoire de Paris) 10, rue Alice Domon et L\'eonie Duquet, 75205 Paris Cedex 13,  France}}\vspace*{0.15cm}
\nopagebreak[3]
\vspace*{-0.20\baselineskip}
\nopagebreak[3]
\address[LAM]{\scriptsize{LAM - Laboratoire d'Astrophysique de Marseille, P\^ole de l'\'Etoile Site de Ch\^ateau-Gombert, rue Fr\'ed\'eric Joliot-Curie 38,  13388 Marseille Cedex 13, France }}\vspace*{0.15cm}
\nopagebreak[3]
\vspace*{-0.20\baselineskip}
\nopagebreak[3]
\address[Bologna-UNI]{\scriptsize{Dipartimento di Fisica dell'Universit\`a, Viale Berti Pichat 6/2, 40127 Bologna, Italy}}\vspace*{0.15cm}
\nopagebreak[3]
\vspace*{-0.20\baselineskip}
\nopagebreak[3]
\address[Bologna]{\scriptsize{INFN - Sezione di Bologna, Viale Berti Pichat 6/2, 40127 Bologna, Italy}}\vspace*{0.15cm}
\nopagebreak[3]
\vspace*{-0.20\baselineskip}
\nopagebreak[3]
\address[Roma-UNI]{\scriptsize{Dipartimento di Fisica dell'Universit\`a La Sapienza, P.le Aldo Moro 2, 00185 Roma, Italy}}\vspace*{0.15cm}
\nopagebreak[3]
\vspace*{-0.20\baselineskip}
\nopagebreak[3]
\address[Rome]{\scriptsize{INFN - Sezione di Roma, P.le Aldo Moro 2, 00185 Roma, Italy}}\vspace*{0.15cm}
\nopagebreak[3]
\vspace*{-0.20\baselineskip}
\nopagebreak[3]
\address[Clermont-Ferrand]{\scriptsize{Clermont Universit\'e, Universit\'e Blaise Pascal, CNRS/IN2P3, Laboratoire de Physique Corpusculaire, BP 10448, 63000 Clermont-Ferrand, France}}\vspace*{0.15cm}
\nopagebreak[3]
\vspace*{-0.20\baselineskip}
\nopagebreak[3]
\address[Pisa-UNI]{\scriptsize{Dipartimento di Fisica dell'Universit\`a, Largo B. Pontecorvo 3, 56127 Pisa, Italy}}\vspace*{0.15cm}
\nopagebreak[3]
\vspace*{-0.20\baselineskip}
\nopagebreak[3]
\address[Pisa]{\scriptsize{INFN - Sezione di Pisa, Largo B. Pontecorvo 3, 56127 Pisa, Italy}}\vspace*{0.15cm}
\nopagebreak[3]
\vspace*{-0.20\baselineskip}
\nopagebreak[3]
\address[INAF]{\scriptsize{INAF-IASF, via P. Gobetti 101, 40129 Bologna, Italy}}\vspace*{0.15cm}
\nopagebreak[3]
\vspace*{-0.20\baselineskip}
\nopagebreak[3]
\address[GEOAZUR]{\scriptsize{G\'eoazur - Universit\'e de Nice Sophia-Antipolis, CNRS/INSU, IRD, Observatoire de la C\^ote d'Azur and Universit\'e  Pierre et Marie Curie, BP 48, 06235 Villefranche-sur-mer, France}}\vspace*{0.15cm}
\nopagebreak[3]
\vspace*{-0.20\baselineskip}
\nopagebreak[3]
\address[LNS]{\scriptsize{INFN - Laboratori Nazionali del Sud (LNS), Via S. Sofia 62, 95123 Catania, Italy}}\vspace*{0.15cm}
\nopagebreak[3]
\vspace*{-0.20\baselineskip}
\nopagebreak[3]
\address[IRFU/SPP]{\scriptsize{Direction des Sciences de la Mati\`ere - Institut de recherche sur les lois fondamentales de l'Univers - Service de Physique des Particules, CEA Saclay, 91191 Gif-sur-Yvette Cedex, France}}\vspace*{0.15cm}
\nopagebreak[3]
\vspace*{-0.20\baselineskip}
\nopagebreak[3]
\address[COM]{\scriptsize{COM - Centre d'Oc\'eanologie de Marseille, CNRS/INSU et Universit\'e de la M\'editerran\'ee, 163 Avenue de Luminy, Case 901, 13288 Marseille Cedex 9, France}}\vspace*{0.15cm}
\nopagebreak[3]
\vspace*{-0.20\baselineskip}
\nopagebreak[3]
\address[UPS]{\scriptsize{Universit\'e Paris-Sud 11 - D\'epartement de Physique, 91403 Orsay Cedex, France}}\vspace*{0.15cm}
\nopagebreak[3]
\vspace*{-0.20\baselineskip}
\nopagebreak[3]
\address[IPHC]{\scriptsize{IPHC - Institut Pluridisciplinaire Hubert Curien - Universit\'e de Strasbourg et CNRS/IN2P3   23 rue du Loess, BP 28,  67037 Strasbourg Cedex 2, France}}\vspace*{0.15cm}
\nopagebreak[3]
\vspace*{-0.20\baselineskip}
\nopagebreak[3]
\address[NIOZ]{\scriptsize{Royal Netherlands Institute for Sea Research (NIOZ), Landsdiep 4,1797 SZ 't Horntje (Texel), The Netherlands}}\vspace*{0.15cm}
\nopagebreak[3]
\vspace*{-0.20\baselineskip}
\nopagebreak[3]
\address[KVI]{\scriptsize{Kernfysisch Versneller Instituut (KVI), University of Groningen, Zernikelaan 25, 9747 AA Groningen, The Netherlands}}\vspace*{0.15cm}
\nopagebreak[3]
\vspace*{-0.20\baselineskip}
\nopagebreak[3]
\address[UU]{\scriptsize{Universiteit Utrecht, Faculteit Betawetenschappen, Princetonplein 5, 3584 CC Utrecht, The Netherlands}}\vspace*{0.15cm}
\nopagebreak[3]
\vspace*{-0.20\baselineskip}
\nopagebreak[3]
\address[UvA]{\scriptsize{Universiteit van Amsterdam, Instituut voor Hoge-Energie Fysika, Science Park 105, 1098 XG Amsterdam, The Netherlands}}\vspace*{0.15cm}
\nopagebreak[3]
\vspace*{-0.20\baselineskip}
\nopagebreak[3]
\address[Catania-UNI]{\scriptsize{Dipartimento di Fisica ed Astronomia dell'Universit\`a, Viale Andrea Doria 6, 95125 Catania, Italy}}\vspace*{0.15cm}
\nopagebreak[3]
\vspace*{-0.20\baselineskip}
\nopagebreak[3]
\address[WIN]{\scriptsize{University of Wisconsin - Madison, 53715, WI, USA}}\vspace*{0.15cm}
\nopagebreak[3]
\vspace*{-0.20\baselineskip}
\nopagebreak[3]
\address[ISS]{\scriptsize{Institute for Space Sciences, R-77125 Bucharest, M\u{a}gurele, Romania }}\vspace*{0.15cm}
\nopagebreak[3]
\vspace*{-0.20\baselineskip}
\nopagebreak[3]
\address[Genova-UNI]{\scriptsize{Dipartimento di Fisica dell'Universit\`a, Via Dodecaneso 33, 16146 Genova, Italy}}\vspace*{0.15cm}
\nopagebreak[3]
\vspace*{-0.20\baselineskip}
\nopagebreak[3]
\address[ITEP]{\scriptsize{ITEP - Institute for Theoretical and Experimental Physics, B. Cheremushkinskaya 25, 117218 Moscow, Russia}}\vspace*{0.15cm}
\nopagebreak[3]
\vspace*{-0.20\baselineskip}
%\end{frontmatter}
%\end{document}

\begin{abstract}
% Text of abstract
The AMADEUS (ANTARES Modules for the Acoustic Detection Under the Sea)
system which is described in this article aims at the investigation of
techniques for acoustic detection of neutrinos in the deep sea. It is
integrated into the ANTARES neutrino telescope in the
Medi\-ter\-ran\-ean Sea.  Its acoustic sensors, installed at water
depths between 2050 and 2300\,m, employ piezo-electric elements for
the broad-band recording of signals with frequencies ranging up to
125\,kHz. The typical sensitivity of the sensors is around
$-$145\,dB\,re\,1V/$\upmu$Pa (including preamplifier).
Completed in May 2008, AMADEUS consists of six ``acoustic clusters'',
each comprising six acoustic sensors that are arranged at
distances of roughly 1\,m from each other. 
Two vertical mechanical structures (so-called lines) of the ANTARES
detector host three acoustic clusters each.  Spacings between the
clusters range from 14.5 to 340\,m.  
Each cluster contains custom-designed electronics boards to
amplify and digitise the acoustic signals from the sensors. 
An on-shore computer cluster 
is used to process and filter the data stream and store the selected events.
The daily volume of recorded data is about 10 GB.
The system is operating continuously and automatically, requiring only
little human intervention.
AMADEUS allows for extensive studies of both transient
signals and ambient noise in the deep sea, as well as signal
correlations on several length scales and localisation of acoustic
point sources.  
Thus the system is excellently suited 
to assess the background conditions for the measurement of the
bipolar pulses expected to originate from neutrino interactions. 

\end{abstract}

\end{frontmatter}

% main text
\section{Introduction}
\label{sec:introduction}

Measuring acoustic pressure pulses in huge underwater acoustic arrays
is a promising approach for the detection of cosmic neutrinos with
energies exceeding 100\,PeV.  The pressure signals are produced by the
particle cascades that evolve when neutrinos interact with nuclei in
water.
The resulting energy deposition in a cylindrical volume of a few
centimetres in radius and several metres in length leads to a local
heating of the medium which is instantaneous with respect to the
hydrodynamic time scales.  This temperature change induces an
expansion or contraction of the medium depending on its volume
expansion coefficient.  According to the thermo-acoustic
model~\cite{bib:Askariyan2,bib:Learned}, the accelerated motion of the
heated volume---a micro-explosion---forms a pressure pulse of bipolar
shape which propagates in the surrounding medium.
Coherent superposition of the elementary sound waves, produced over the
volume of the energy deposition, leads to a propagation within a flat
disk-like volume (often referred to as {\em pancake})
in the direction perpendicular to the axis of the particle cascade.
After propagating several hundreds of metres in sea water, the pulse
has a characteristic frequency spectrum that is expected to peak
around 10\,kHz~\cite{bib:Sim_Acorne,bib:Sim_Acorne2,bib:Bertin_Niess}.
Given the strongly anisotropic propagation pattern of the sound waves,
the details of the pressure pulse, namely its amplitude, asymmetry and
frequency spectrum, depend on the distance and angular position of the
observer with respect to the particle cascade induced by the neutrino
interaction~\cite{bib:Sim_Acorne}.
Besides sea water, which is the medium under investigation in the case
of the AMADEUS (ANTARES Modules for the Acoustic Detection Under the
Sea) project, ice~\cite{bib:spats} and fresh water~\cite{bib:baikal}
are investigated as media for acoustic detection of neutrinos.
Studies in sea water are also pursued by other groups using military
arrays of hydrophones (i.e. underwater
microphones)~\cite{bib:saund,bib:acorne} or exploiting other existing
deep sea infrastructures~\cite{bib:noise_ONDE}.

Two major advantages over an optical neutrino telescope motivate
studying acoustic detection.  First, the attenuation length in sea water
is about 5\,km (1\,km) for 10\,kHz (20\,kHz) signals.  This
is one to two orders of magnitude larger than for visible light
with a maximum attenuation length of about 60\,m.
The second advantage is the more compact sensor design and simpler readout
electronics for acoustic measurements. 
Since on the other hand the speed of sound\footnote{The speed of sound
  in sea water depends on temperature, salinity and pressure,
  i.e. depth. A good guideline value for the speed of sound at the
  location of AMADEUS is 1500\,m/s.}  is small compared to the speed
of light, coincidence windows between two spatially separated sensors
are correspondingly large. Furthermore, the signal amplitude is
relatively small compared to the acoustic background in the sea,
resulting in a high trigger rate at the level of individual sensors
and making the implementation of efficient online data reduction
techniques essential.  To reduce the required processing time without
sacrificing the advantages given by the large attenuation length, the
concept of spatially separated clusters of acoustic sensors is used in
AMADEUS.  Online data filtering is then predominantly applied to the
closely arranged sensors within a cluster.

The AMADEUS project was conceived to perform a feasibility study for a
potential future large-scale acoustic neutrino detector. For this purpose, 
a dedicated array of acoustic sensors was integrated into the
ANTARES neutrino telescope~\cite{bib:ANTARES-paper, bib:ANTARES-line1}. 
In the context of AMADEUS, the following aims are being pursued:
\begin{itemize}
\item
Long-term background investigations 
(levels of ambient noise, spatial and temporal distributions of
sources, rate of neutrino-like signals);

\item
Investigation of spatial correlations for 
transient signals and for persistent background
on different length scales;
\item
Development and tests of data filter and reconstruction algorithms;
\item
Investigation of different types of acoustic sensors and sensing methods;
\item
Studies of hybrid (acoustic and optical) detection methods.
\end{itemize}

In particular the knowledge of the rate and correlation length of
neutrino-like acoustic background events
is a prerequisite for estimating the sensitivity of a future acoustic
neutrino detector.

The focus of this paper is the AMADEUS system within the ANTARES
detector.
In Section~\ref{sec:amadeus_overview}, 
an overview of the system is given, with particular emphasis on 
its integration into the ANTARES detector.
In Section~\ref{sec:syscomponents}, the system components are
described and in Section~\ref{sec:sysperformance} the system
performance is discussed.  The characteristic features of the
AMADEUS system are mainly determined by the acoustic
sensors and the custom-designed electronics board, which performs
the off-shore processing of the analogue signals from the acoustic
sensors.  These two components are discussed in detail in Subsections
\ref{sec:acousensors} and \ref{sec:acouadc-board}.

\section{Overview of the AMADEUS System}
\label{sec:amadeus_overview}

\subsection{The ANTARES Detector and its Sub-system AMADEUS}
\label{sec:amadeus_part_antares}

\begin{figure}[ht]
\centering
\includegraphics[width=10.0cm]{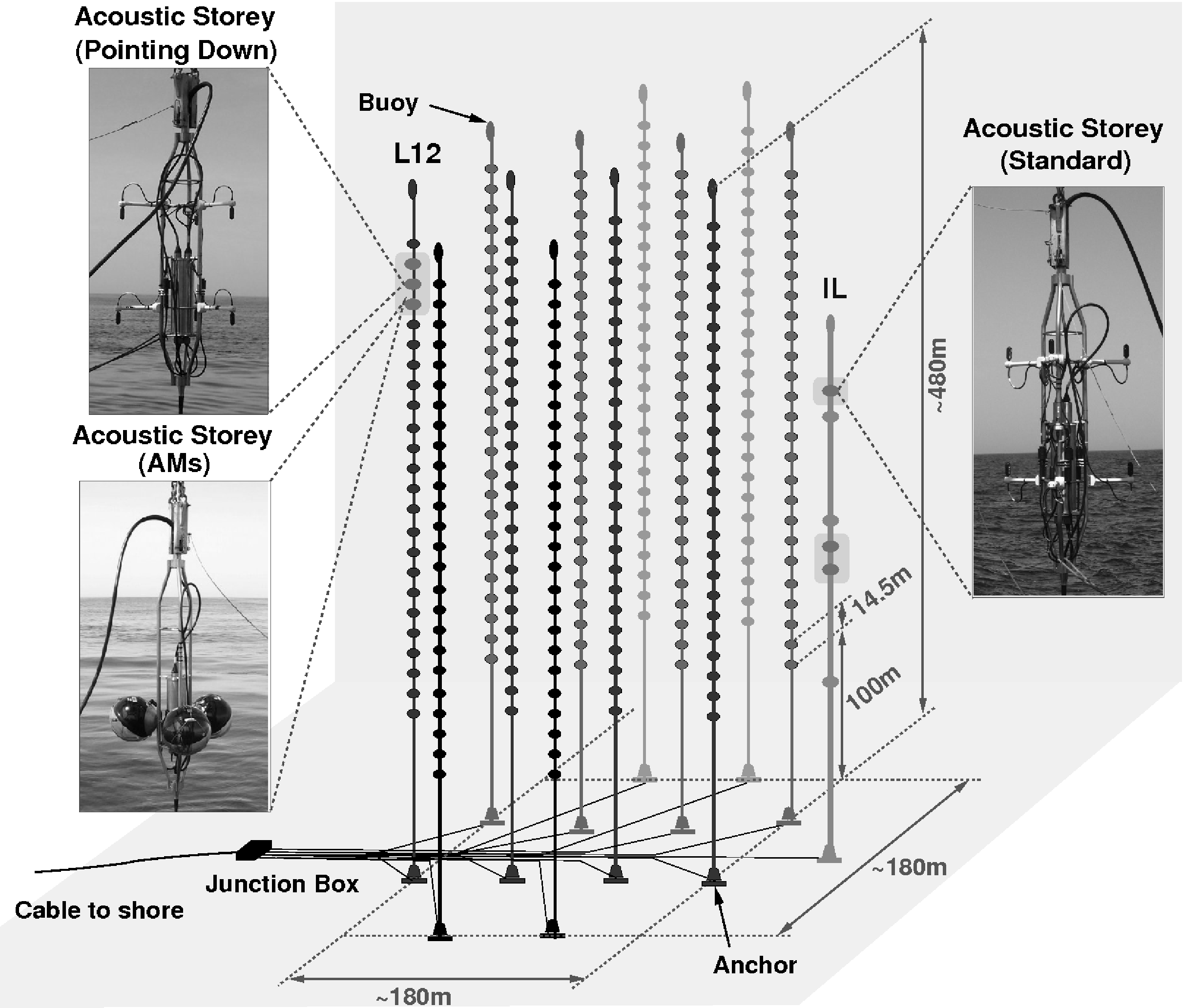}
\caption{A sketch of the ANTARES detector. 
The six acoustic storeys are highlighted and their three different setups
are shown (see text for details). 
L12 and IL denote the 12th detection line and the Instrumentation Line, 
respectively.}
\label{fig:ANTARES_schematic_all_storeys}
\end{figure}

AMADEUS is integrated into the ANTARES neutrino
telescope~\cite{bib:ANTARES-paper, bib:ANTARES-line1}, which was
designed to detect neutrinos by measuring the Cherenkov light emitted
along the tracks of relativistic secondary muons generated in neutrino
interactions.
A sketch of the detector, with the AMADEUS modules highlighted, is
shown in Figure~\ref{fig:ANTARES_schematic_all_storeys}.  The detector
is located in the Mediterranean Sea at a water depth of about 2500\,m,
about 40\,km south of the town of Toulon on the French coast and was
completed in May 2008.
It comprises 12 vertical structures, the {\em detection lines}. 
Each detection line
holds up to 25 {\em storeys} that are arranged at equal distances of 14.5\,m
along the line, starting at about 100\,m above the sea
bed and interlinked by electro-optical cables.  A standard
storey consists of a titanium support structure, holding three {\em
  Optical Modules}\cite{bib:OMs} 
(each one consisting of a photomultiplier tube (PMT) inside
a water-tight pressure-resistant glass sphere) and one {\em Local
  Control Module (LCM).}  
The LCM consists of 
a cylindrical titanium container
and the off-shore electronics within that container
(see Subsection~\ref{sec:offshore}).

A 13th line, called {\em Instrumentation Line (IL)}, is equipped with
instruments for monitoring the environment. It holds six storeys.
For two pairs of consecutive storeys in the IL, the vertical distance
is increased to 80\,m.
Each line is fixed on the sea floor by an anchor equipped with
electronics and held taut by an immersed buoy.  An interlink cable
connects each line to the {\em Junction Box} from where the main
electro-optical cable provides the connection to the shore station.

The ANTARES lines are free to swing and twist in the undersea current.
In order to determine the positions of the storey with a precision of
about 20\,cm---which is necessary to achieve the required pointing
precision for neutrino astronomy---the detector is equipped with an
acoustic positioning system~\cite{bib:antares-pos}.  The system
employs an acoustic transceiver at the anchor of each line and four
autonomous transponders positioned around the 13 lines. Along each
detection line, five positioning hydrophones
receive the signals emitted by the transceivers.  By performing
multiple time delay measurements and using these to triangulate the
positions of the individual hydrophones,
the line shapes can be reconstructed relative to the positions of the
emitters. Currently, the sequence of positioning emissions is repeated
every 2 minutes.

In AMADEUS, acoustic sensing is integrated in the form of {\em
  acoustic storeys} that are modified versions of standard ANTARES
storeys, in which the Optical Modules are replaced by custom-designed
acoustic sensors.  Dedicated electronics is used for the digitisation
and pre-processing of the analogue signals.  The acoustic storeys are
equivalent to the acoustic clusters introduced in
Section~\ref{sec:introduction}.

The AMADEUS system comprises a total of six acoustic storeys: three on
the IL, which started data taking in December 2007, and three on the
12th detection line (Line 12), which was connected to shore in May
2008.  AMADEUS is now fully functional and routinely taking data with
34 sensors. Two out of 36 hydrophones became inoperational during
their deployment.  In both cases, the defect was due to
pressurisation.

The acoustic storeys on the IL are located at 180\,m, 195\,m, and
305\,m above the sea floor.  On Line 12, which is anchored at a
horizontal distance of about 240\,m from the IL, the acoustic storeys
are positioned at heights of 380\,m, 395\,m, and 410\,m above the sea
floor.  With this setup, the maximum distance between two acoustic
storeys is 340\,m.
AMADEUS hence covers three length scales: spacings of the order of
1\,m between sensors within a storey (i.e. an 
acoustic cluster); intermediate
distances of 14.5\,m between adjacent acoustic storeys within a
line; and large scales from about 100\,m vertical distance on the IL
up to 340\,m between storeys on different lines.
The sensors within a cluster allow for efficient triggering of
transient signals and for direction reconstruction.  The combination
of the direction information from different acoustic storeys yields
(after verifying the consistency of the signal arrival times at the
respective storeys) the position of an acoustic
source~\cite{bib:Richardt_reco}.
The AMADEUS system includes
time synchronisation and a continuously operating data acquisition
setup and is in principle scalable to a large-volume detector.

\subsection{Acoustic Storeys}
\label{sec:acou_storey}
\begin{figure}[ht]
\centering
\includegraphics[height=8.0cm]{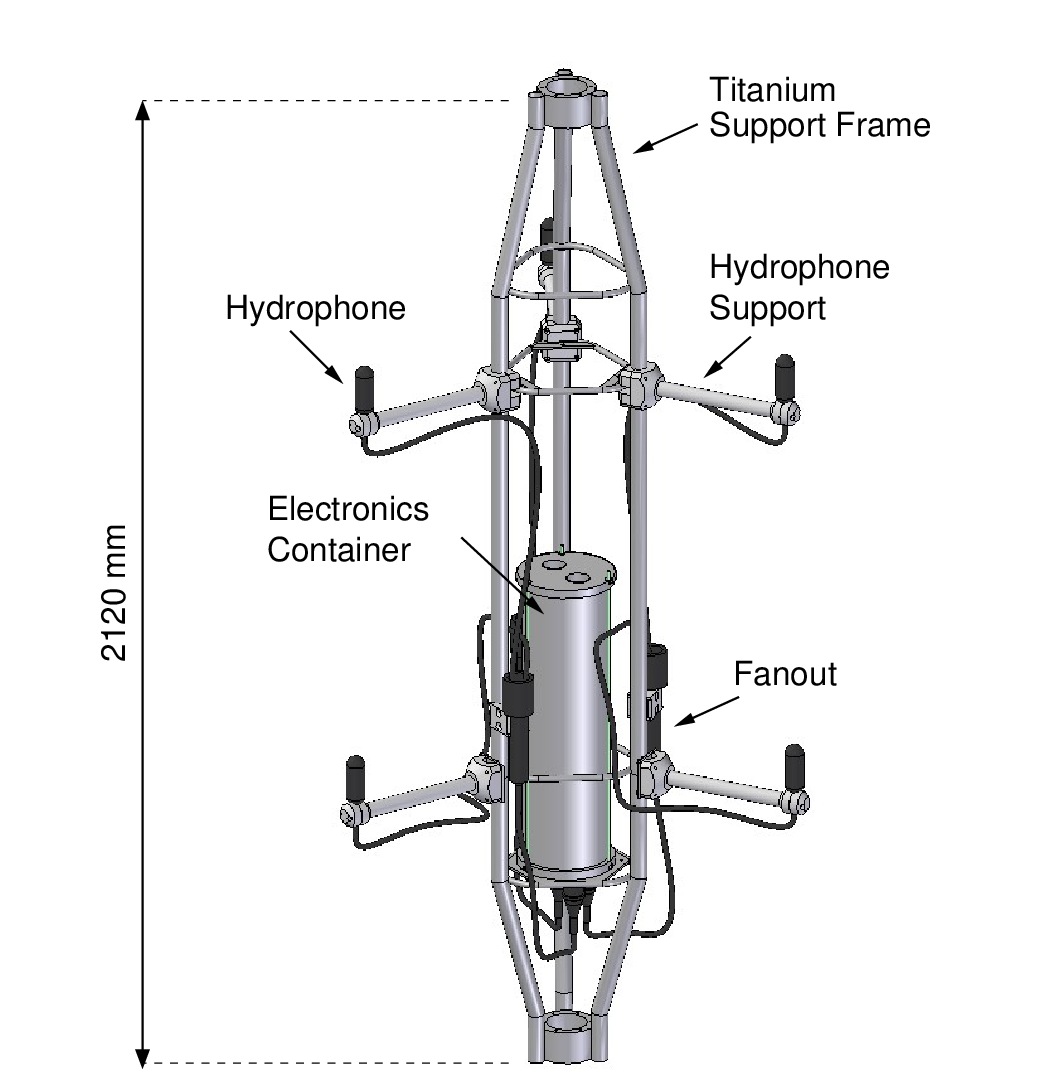}
\caption{
Drawing of a standard acoustic storey,  or acoustic cluster, 
with hydrophones.
\label{fig:acou_storey_drawing}
}
\end{figure}

Two types of sensing devices are used in AMADEUS: hydrophones and {\em
  Acoustic Modules} (AMs).  The sensing principle is in both cases
based on the piezo-electric effect and is discussed in
Subsection~\ref{sec:acousensors}.  
For the hydrophones, the piezo elements are coated in polyurethane,
whereas for the AMs they are glued to the inside of
standard glass spheres which are normally used for Optical Modules.
Figure~\ref{fig:acou_storey_drawing} shows the design of a standard
acoustic storey with hydrophones.

\begin{figure}[ht]
\centering
\subfigure[]{
\includegraphics[height=7.2cm]{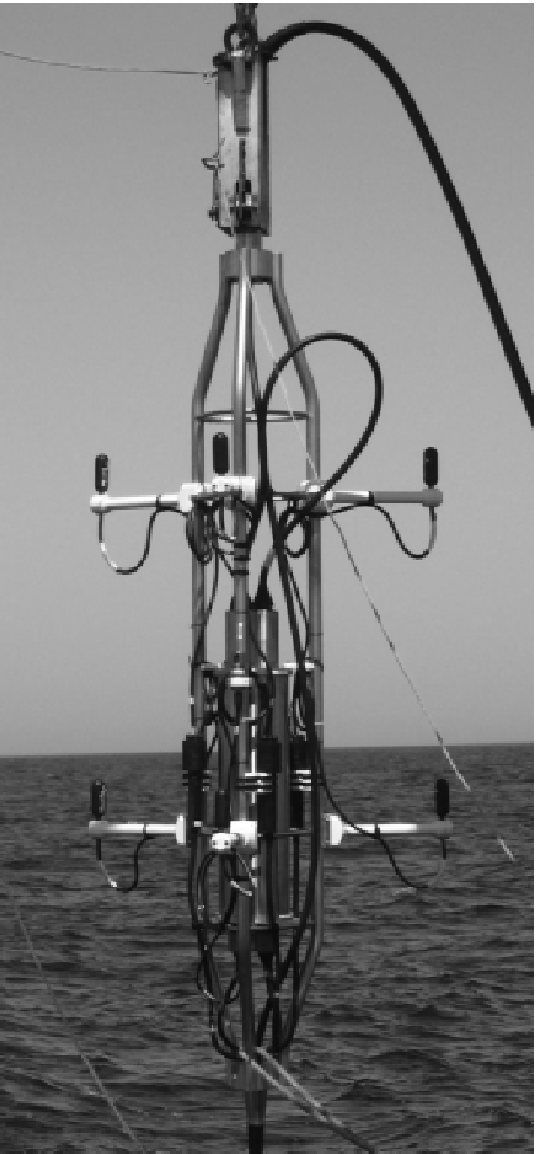}
\label{subfig:storey_standard}
}
\hspace{0.75mm}
\subfigure[]{
\includegraphics[height=7.2cm]{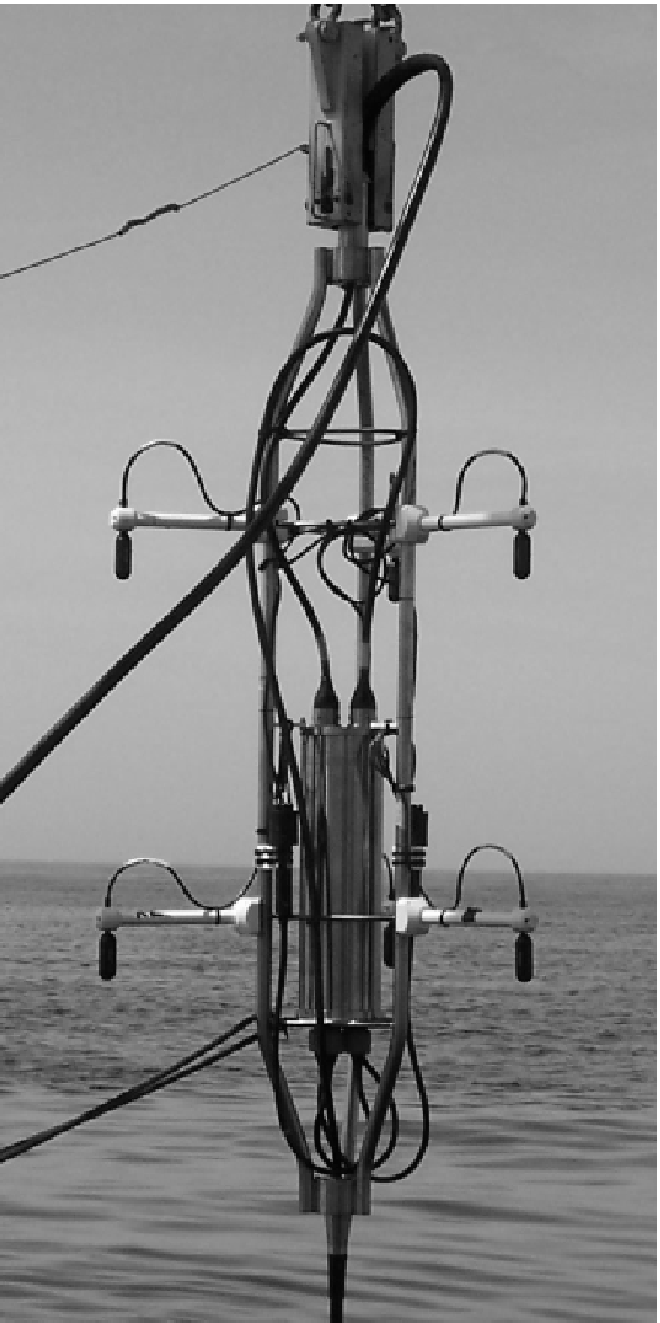}
\label{subfig:storey21}
}
\hspace{0.75mm}
\subfigure[]{
\includegraphics[height=7.2cm]{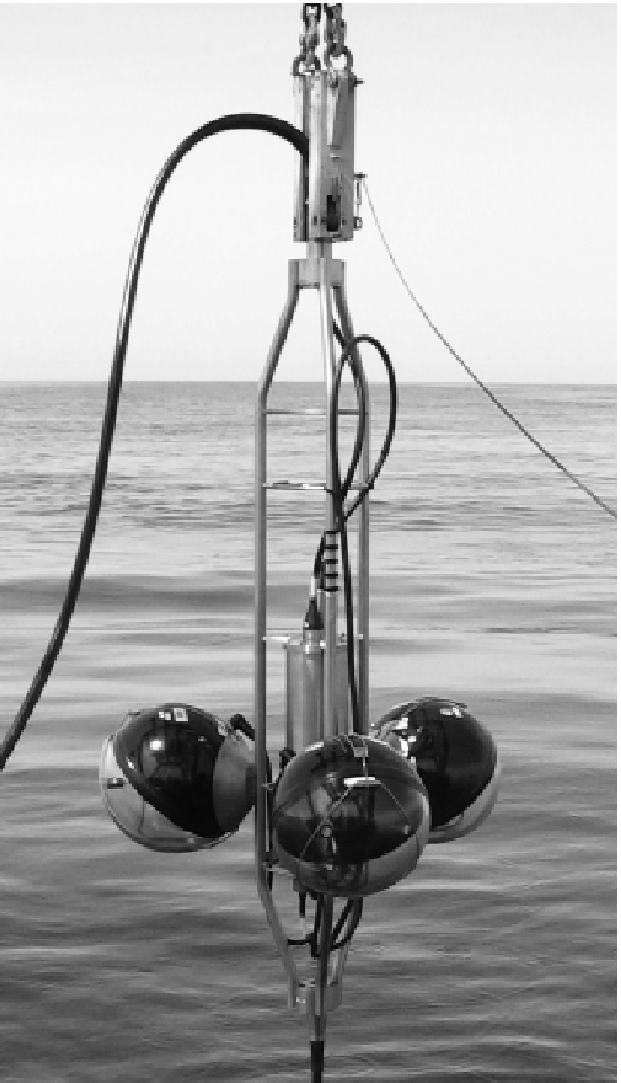}
\label{subfig:storey22}
}
\caption{Photographs of three different storeys of the AMADEUS
  system during their deployment: 
  \subref{subfig:storey_standard}
  A standard storey, equipped with hydrophones pointing up;
  \subref{subfig:storey21} the central acoustic storey on Line 12 with
  the hydrophones pointing down;  
  \subref{subfig:storey22} the
  lowermost acoustic storey on Line 12 equipped with Acoustic Modules.
}
\label{fig:antares_storey_acou}
\end{figure}

Figure~\ref{fig:antares_storey_acou} shows the three different designs
of acoustic storeys installed in AMADEUS.  The acoustic storeys on the
IL house hydrophones only, whereas the lowermost acoustic storey of
Line 12 holds AMs.  The hydrophones are mounted to point upwards,
except for the central acoustic storey of Line 12, where they point
downwards.  The sensitivity of the hydrophones is largely reduced at
their cable junctions and therefore shows a strong dependence on the
polar angle.  The different configurations allow for investigating the
anisotropy of ambient noise, which is expected to originate mainly
from the sea surface.

Three of the five storeys holding hydrophones are equipped with
commercial models, dubbed ``HTI hydrophones''\footnote{Custom produced
  by High Tech Inc (HTI) in Gulfport, MS (USA).}, and the other two
with hydrophones, described in detail in
Subsection~\ref{sec:acousensors}, developed and produced at the
Erlangen Centre for Astroparticle Physics (ECAP).

\subsection{Design Principles}

A fundamental design guideline for the AMADEUS system has been to use
existing ANTARES hardware and software as much as possible.  This eases
the operation of the system within the environment of the ANTARES
neutrino telescope; at the same time, the design efforts were kept to a
minimum and new quality assurance and control measures had to be
introduced only for the additional components. These were subjected to
intensive testing procedures, in particular in view of the hostile
environment due to the high water pressure of up to 240 bar and the
salinity of the water.

In order to integrate the AMADEUS system into the ANTARES neutrino
telescope, design and development efforts in the following basic areas
were necessary:
\begin{itemize}
\item The development of acoustic sensing devices that replace the
  Optical Modules of standard ANTARES storeys and of the cables to route
  the signals into the electronics container;
\item The
development of an off-shore acoustic digitisation and pre-processing board; 
\item The setup of an on-shore server cluster for the online processing of the
acoustic data and the development of the online software;
\item The development of offline reconstruction and simulation software. 
\end{itemize}

Six acoustic sensors per storey were implemented. This number was the
maximum compatible with the design of the LCM and the bandwidth of
data transmission to shore.  Furthermore, the acoustic storeys were
designed such that their size did not exceed the size of the standard
ANTARES storeys in radial dimension, hence assuring compatibility with
the deployment procedure of the ANTARES lines.

\subsection{The AMADEUS-0 Test Apparatus}
In March 2005, a full-scale mechanical prototype line for the ANTARES
detector was deployed and subsequently recovered~\cite{bib:line0}. 
This line, dubbed {\em Line 0}, contained no photomultipliers and no
readout electronics.  Instead, an autonomous data logging system and
shore-based optical time-domain reflectometry were used to record the
status of the setup.

Line 0 provided a well-suited environment to study the properties of
the acoustic sensors in situ at a time when the readout electronics
for AMADEUS was still in the planning phase and the piezo-preamplifier
setup in the design phase.  For this purpose, an autonomous system
within a standard LCM container, the {\em AMADEUS-0} device, was
integrated into Line 0.  It recorded acoustic signals at the ANTARES
site using five piezo sensors with custom-designed preamplifiers with
an overall sensitivity of about $-$120\,dB\,re\,1V/$\upmu$Pa in the
range from 5 to 50\,kHz, glued to the inside of the LCM container.  A
battery-powered readout and data logging system was devised and
implemented using commercially available components.  The system was
further equipped with a timing mechanism to record data over two
pre-defined periods:
The first one lasted for about 10 hours and included the deployment of
the line. During this period, a total of 2:45 hours of data were
recorded over several intervals.  In the second period, with the line
installed on the sea floor, 1:45 hours of data were taken over a
period of 3:30 hours until the battery power was exhausted.

\begin{figure}[ht]
\centering
\includegraphics[height=7.0cm]{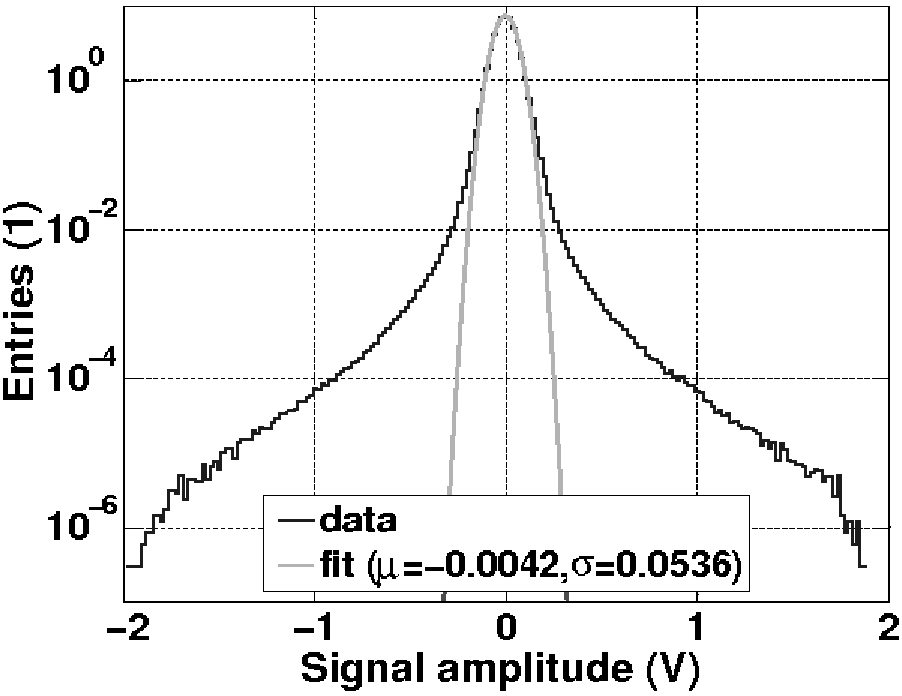}
\caption{
\label{fig:/AMA0_amphist}
Normalised distribution of signal amplitudes for all data recorded
with the AMADEUS-0 device.  High amplitude ($>$2\,V) signals,
saturating the readout electronics, have been removed. A gaussian fit
to the data yields mean $\mu$ and standard deviation $\sigma$.  }
\end{figure}

The analysis of the data~\cite{bib:Deffner_diplom} provided valuable
information for the design of the AMADEUS system. In particular, the
level of the recorded noise allowed for tuning the sensitivity and
frequency response of the preamplifiers and amplifiers.
A filtered amplitude distribution is shown in
Figure~\ref{fig:/AMA0_amphist}, where signals saturating the readout
electronics have been removed. The gaussian fit shown in the figure is
a measure of the combined ambient noise of the deep sea and inherent
noise of the system, while the excess of data is due to transient
signals. This shows that the sensitivity of the system is well matched
to record background noise while at the same time allowing for a wide
dynamic range of transient signals. A comparable overall sensitivity
was hence chosen for the AMADEUS setup. The design of the commercial
readout electronics proved to be not suitable in terms of long-term
stability and the response to signals that saturated the readout
electronics.  This experience was returned to the design of the
AMADEUS readout electronics, which will be described in
Section~\ref{sec:acouadc-board}.

\section{System Components}
\label{sec:syscomponents}

\subsection{The Acoustic Sensors}
\label{sec:acousensors}

The fundamental components of both the hydrophones and the AMs, 
collectively referred to as acoustic sensors,
are piezo-electrical ceramic elements, converting pressure
waves into voltage signals~\cite{bib:hoessl2006}, and preamplifiers.
In this subsection, the hydrophones, the AM sensors, and 
the calibration of their sensitivity will be discussed.

\subsubsection{Hydrophones}
\begin{figure}[ht]
\centering
\includegraphics[width=8.5cm]{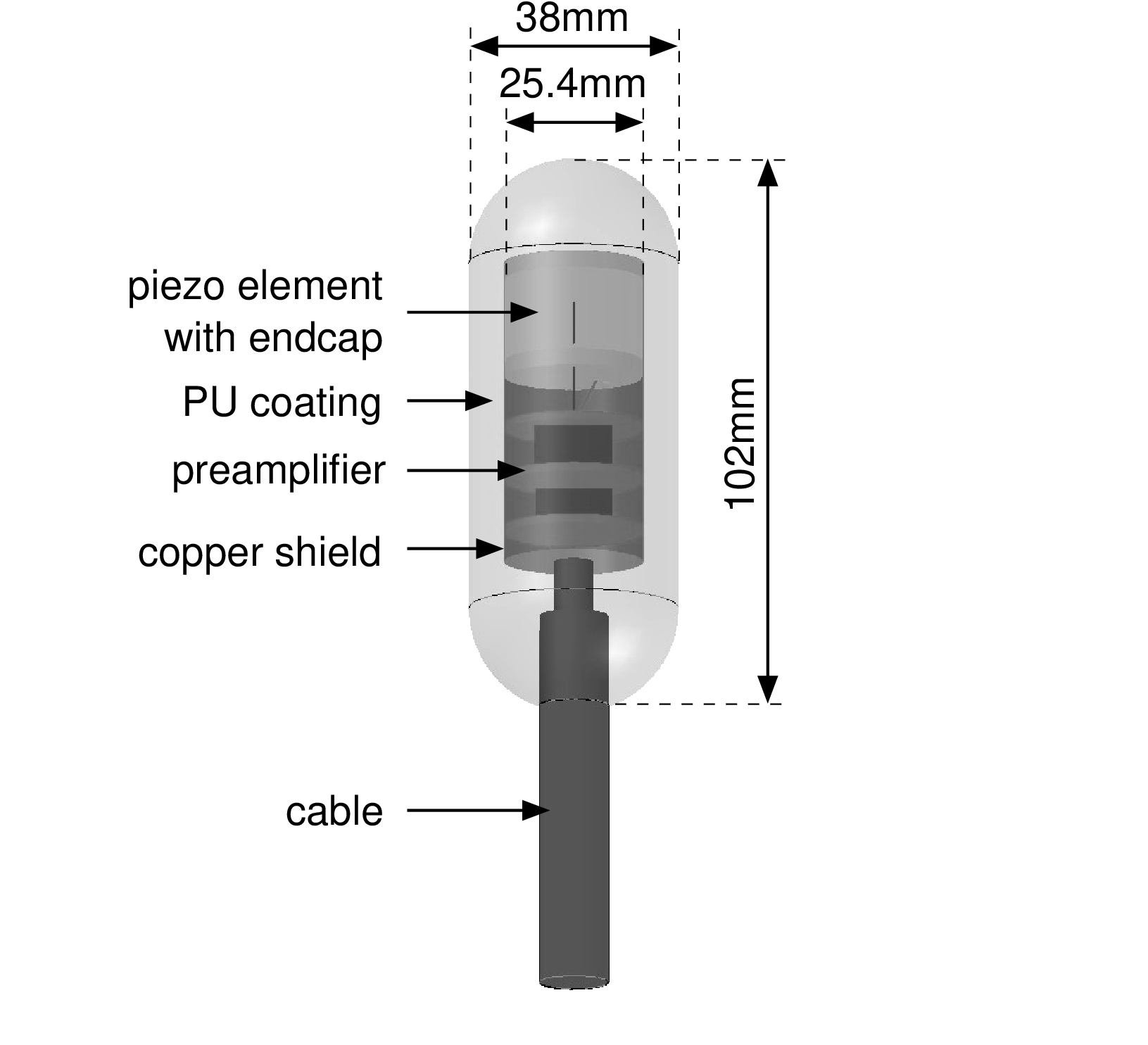}
\hspace{1mm}
\caption{Schematic drawing of an ECAP hydrophone. Piezo element
  and preamplifier 
(consisting of three circular circuit boards, interconnected
by pin connectors) 
are moulded into polyurethane (PU).
\label{fig:lti_hydro-scheme}
}
\end{figure}

A schematic drawing of an ECAP hydrophone is shown in
Figure~\ref{fig:lti_hydro-scheme}.
For these hydrophones\footnote{For the commercial hydrophones, details
  were not disclosed by the manufacturer, but the main design is
  similar to the one described here.}, two-stage preamplifiers were
used: adapted to the capacitive nature of the piezo elements and the
low induced voltages, the first preamplifier stage is charge
integrating while the second one is amplifying the output voltage of
the first stage.  The shape of the ceramics is that of a hollow
cylinder.

Due to hardware constraints of the electronics container, the only
voltage available for the operation of the preamplifiers
was 6.0\,V. In order to minimise electronic noise, 
the preamplifiers were designed for that voltage rather than
employing DC/DC converters to obtain the 12.0\,V supply typically used.

The piezo elements and preamplifiers of the hydrophones are coated in 
polyurethane.
Plastic endcaps prevent the material from pouring into the hollow part
of the piezo cylinder during the moulding procedure. 
 The ECAP as well as the HTI hydrophones have a diameter of 38\,mm and
a length (from the cable junction to the opposite end) of 102\,mm. 

The equivalent inherent noise level in the frequency range from 1 to
50\,kHz is about 13\,mPa for the ECAP hydrophones and about 5.4\,mPa
for the HTI hydrophones. This compares to 6.2\,mPa of the lowest
expected ambient noise level in the same frequency band for a
completely calm sea~\cite{bib:Graf_PhD_2008},
referred to as {\em sea state 0}~\cite{urick2}. 

At the ANTARES site, the hydrophones are subject to an external
pressure of 210 to 240 bar, depending on the depth at which they are
installed.  Prior to deployment, each hydrophone was pressure-tested
in accordance with the standard ANTARES procedure, i.e. the pressure
was ramped up to 310 bar at 12 bar per minute, held there for two
hours and then ramped down again at 12 bar per minute.

\subsubsection{Acoustic Modules}
\begin{figure}[ht]
\centering
\subfigure[]{
\includegraphics[height=5.5cm]{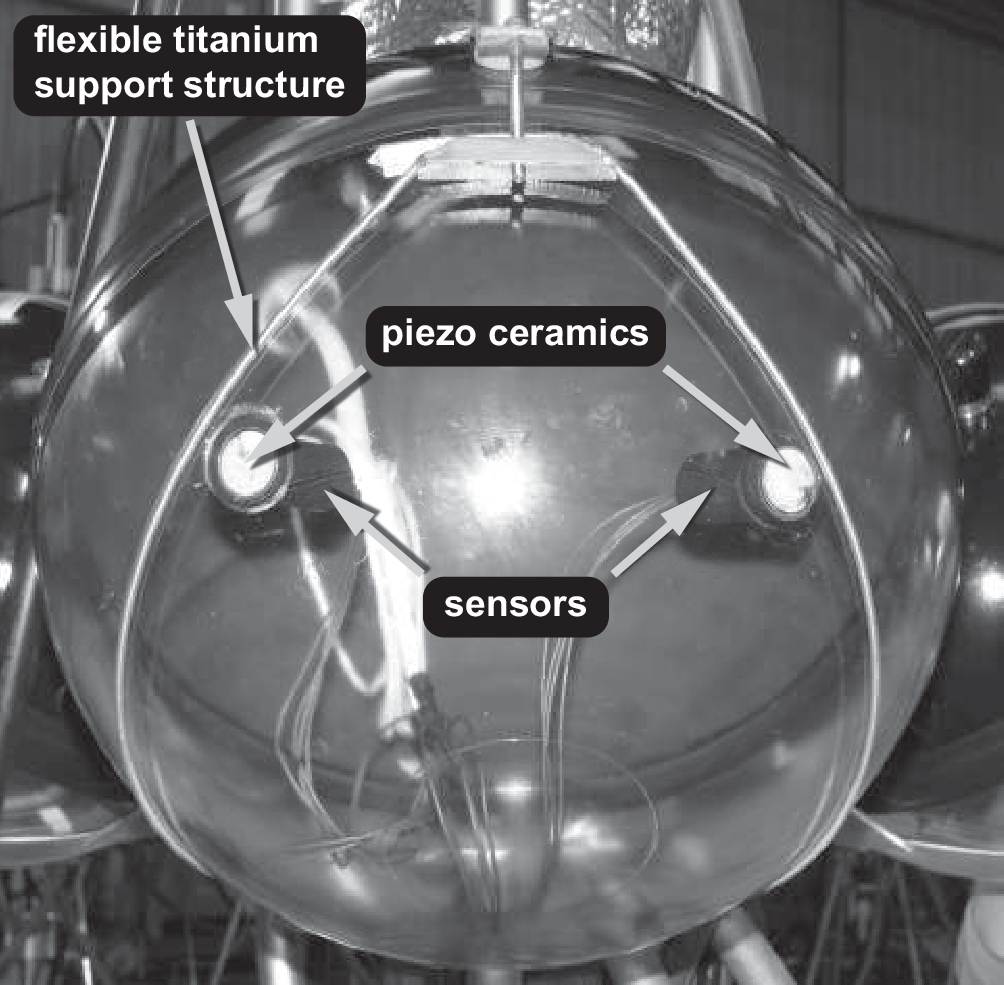}
\label{subfig:AM_photo}
}
\hspace{5mm}
\subfigure[]{
\includegraphics[height=5.5cm]{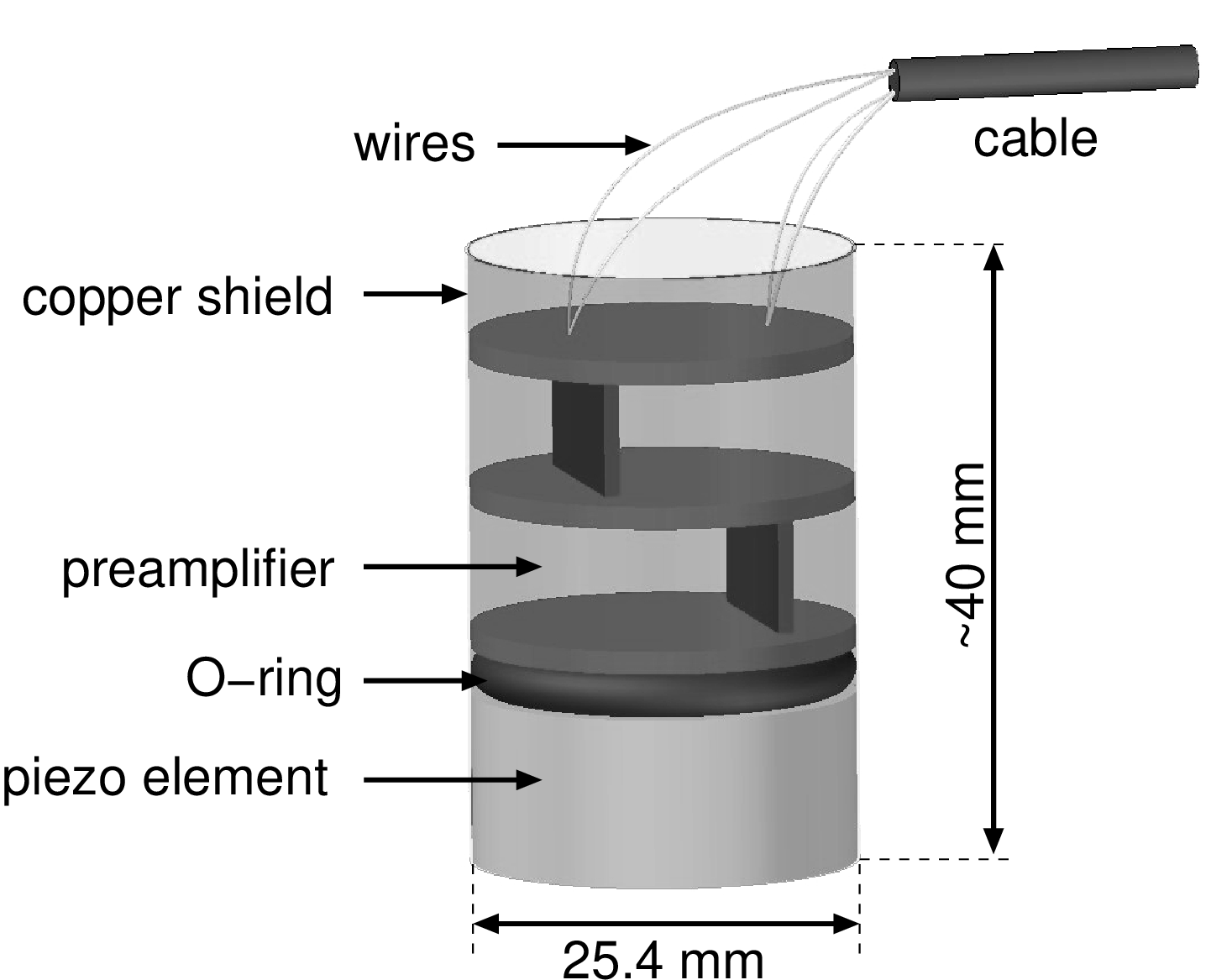}
\label{subfig:AM_sensor}
} \newline
\subfigure[]{
\includegraphics[height=7cm]{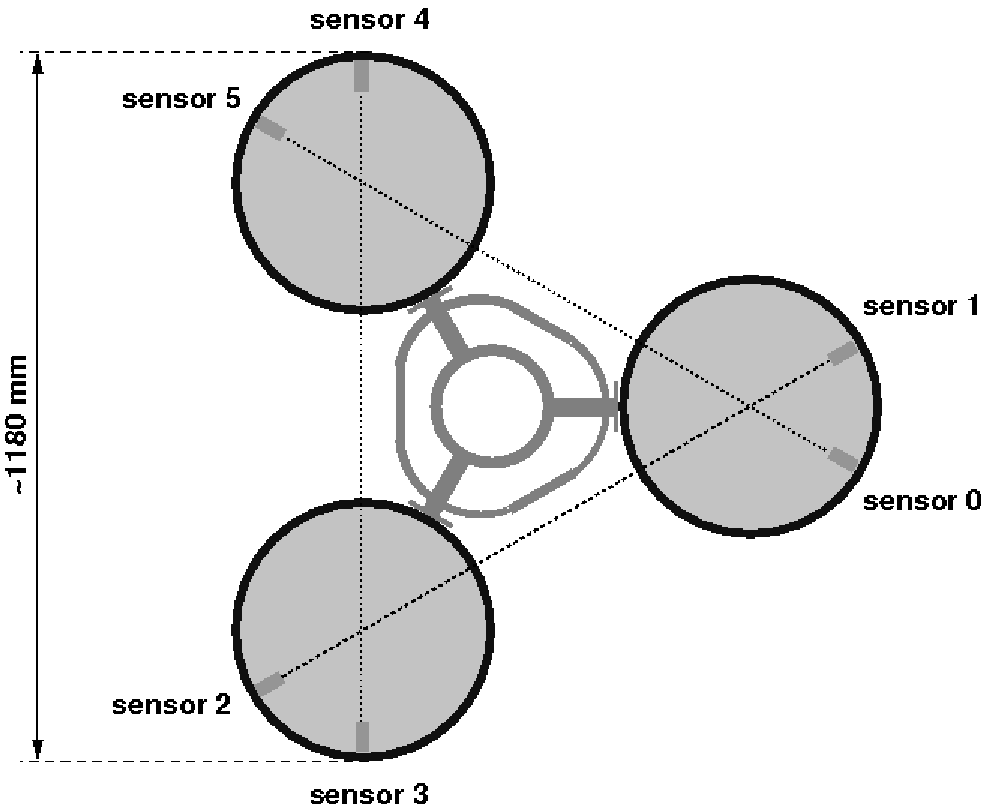}
\label{subfig:AM_schematic_gray}
}
\caption{ \subref{subfig:AM_photo} Photograph of an Acoustic Module
  (AM) before deployment; \subref{subfig:AM_sensor} schematic drawing
  of an AM sensor; \subref{subfig:AM_schematic_gray} horizontal
  cross-section of an acoustic storey holding Acoustic Modules in the
  plane of the sensors.  
  The dotted lines are collinear with the
  longitudinal axes of the sensors and indicate the arrangement of the
  sensor within the storey. The lines intersect at angles of 60$^\circ$ at
  the centres of the glass spheres.
\label{fig:AM_gray}
}
\end{figure}

For the AMs, the same preamplifiers are used as for the ECAP
hydrophones. The piezo elements have the same outer diameter but are
solid cylinders in case of the AMs.  Two sensors are glued to the
inside of each sphere.  This design was motivated by the idea to
operate the piezo elements at low pressure and also to investigate an
option for acoustic sensing that can be integrated together with a PMT
in the same glass sphere.  In order to assure a good acoustic
coupling, the space between the curved sphere and the flat end of the
piezo sensor of the AMs was filled with epoxy.
A photograph of an Acoustic Module and a schematic drawing of the 
sensors glued to the inside of the glass sphere are shown in
Figures~\ref{subfig:AM_photo} and \ref{subfig:AM_sensor}, respectively.

In order to obtain a  2$\pi$ azimuthal coverage, 
the six sensors are distributed over the three AMs of the storey 
within the horizontal plane defined by the three centres of the spheres
as shown in 
Figure~\ref{subfig:AM_schematic_gray}.
The spheres have outer diameters of 432\,mm.

\begin{figure}[tbh]
\centering

\includegraphics[width=12.0cm]{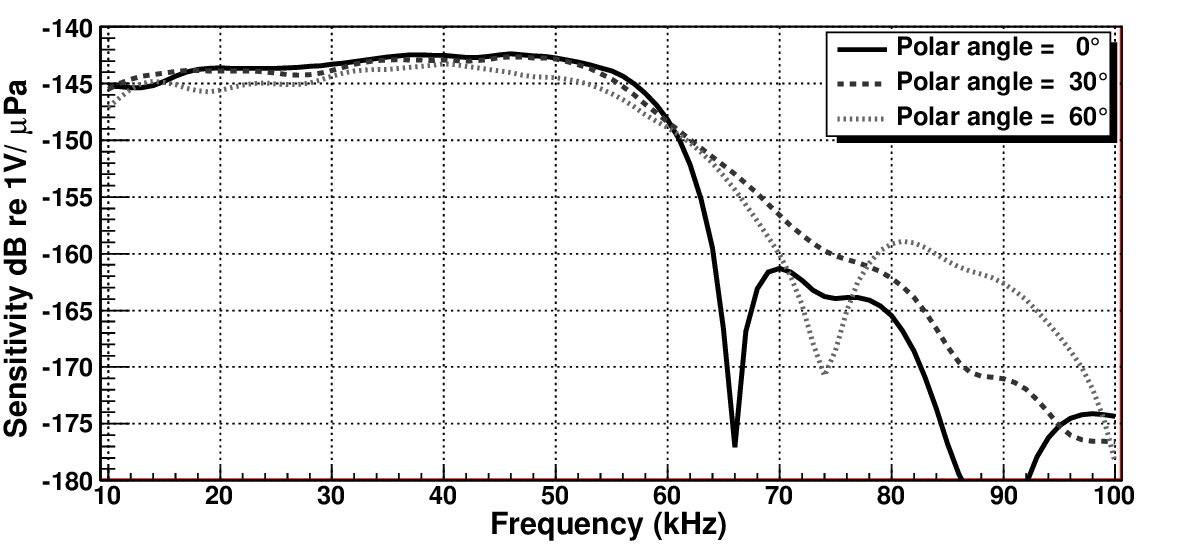}
%\vspace{-1mm}

\includegraphics[width=12.0cm]{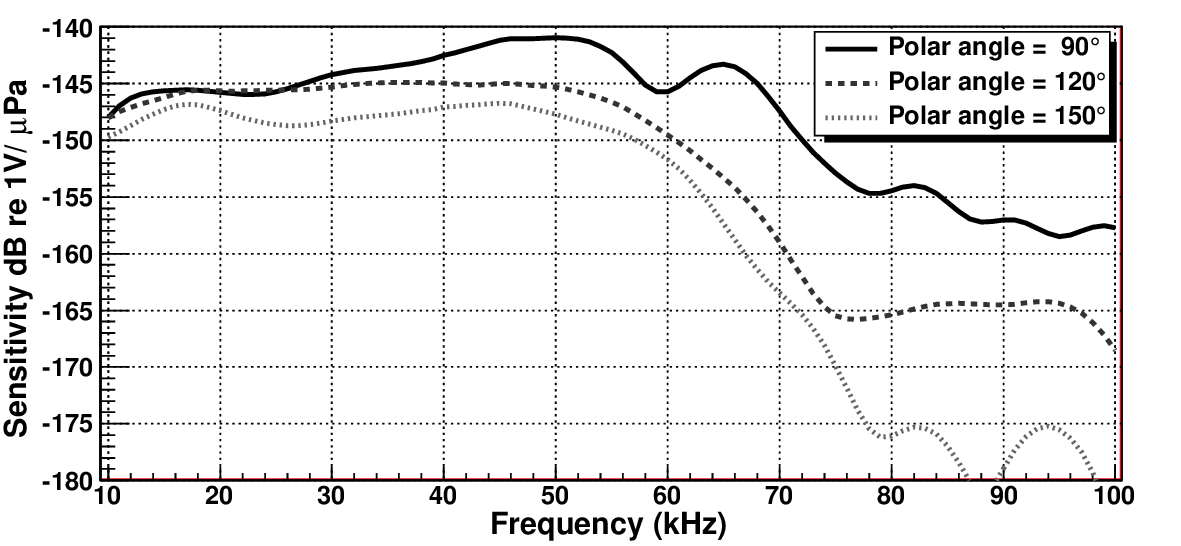}
%\vspace{-1mm}

\caption{
Typical sensitivity of an HTI hydrophone 
as a function of frequency for different polar angles, averaged over the
azimuthal angle. Systematic uncertainties below 50\,kHz are 2 to 3\,dB. 
}
\label{fig:hydrophone_sensitivity}
\end{figure}

\begin{figure}[tbh]
\centering

\includegraphics[width=12.0cm]{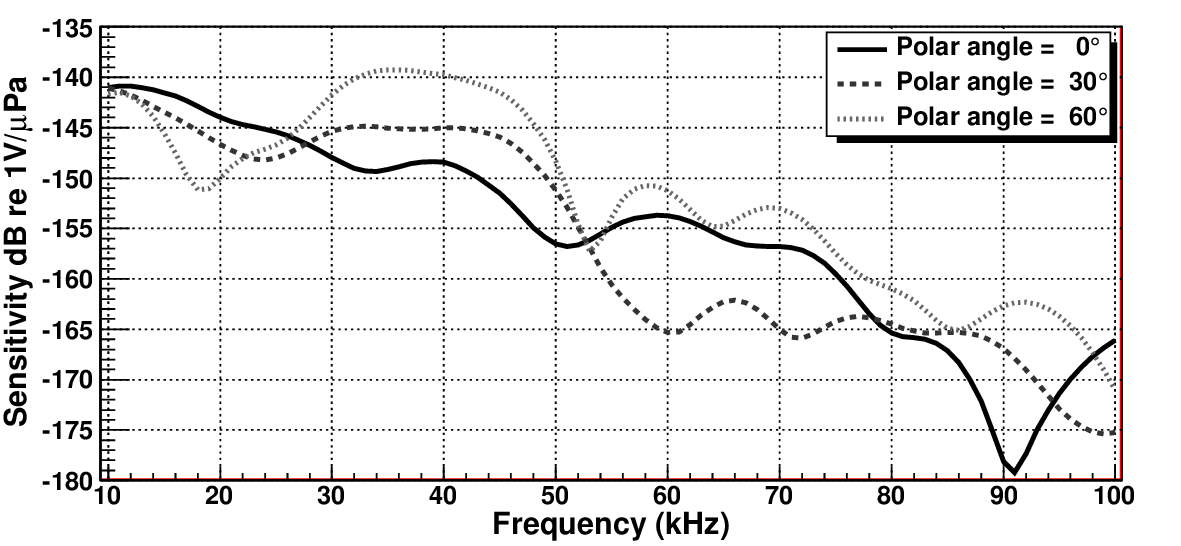}
%\vspace{-1mm}

\includegraphics[width=12.0cm]{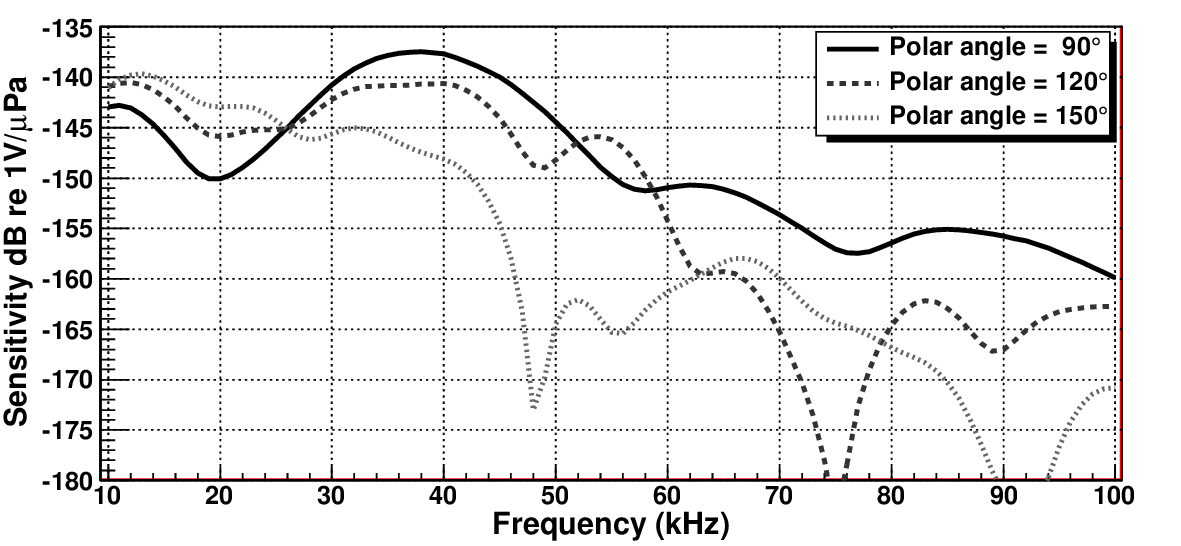}
%\vspace{-1mm}

\caption{
Typical sensitivity of an ECAP hydrophone 
as a function of frequency for different polar angles, averaged over the
azimuthal angle. Systematic uncertainties below 50\,kHz are 2 to 3\,dB. 
}
\label{fig:hydrophone_sensitivity_ECAP}
\end{figure}

\subsubsection{Calibration}

All sensors are tuned to have a low noise level and to be sensitive
over the frequency range from 1 to 50\,kHz with a typical sensitivity
around $-$145\,dB\,re\,1V/$\upmu$Pa (including preamplifier).  The
sensitivities of all sensors as a function of frequency, polar angle
and azimuthal angle were measured before deployment in a water tank,
using a calibrated emitter~\cite{bib:naumann_phd}.  
The analysis was restricted to frequencies above 10\,kHz.
Towards lower frequencies, measurements become increasingly less
significant. This is due to the quadratic frequency dependence of the
emitter's transmit voltage response and to the increasingly adverse
effect of reflections for increasing wavelengths.
In accordance with the expected behaviour of the 
piezo elements, the sensitivity is assumed to be constant 
below 10\,kHz.

The sensitivity of one of the commercial hydrophones
is shown in Figure~\ref{fig:hydrophone_sensitivity} as a function of
frequency for different polar angles.  
For frequencies below 50\,kHz, the sensitivity decreases once the
polar angle approaches 180$^\circ$, which defines the direction at
which the cable is attached to the hydrophone. The beginning of this
trend can be seen for the polar angle of 150$^\circ$.

The sensitivity as a function of the azimuthal angle for a given
frequency is essentially flat at the 3\,dB level for all hydrophones.
The sensitivity as a function of polar angle and frequency shows
deviations of less than 2\,dB between different HTI hydrophones in the
frequency range from 10 to 50\,kHz. The deviations for the hydrophones
produced at ECAP are at a level of 3 to 4\,dB.
The sensitivity of an ECAP hydrophone
is shown in Figure~\ref{fig:hydrophone_sensitivity_ECAP}.
Compared to the HTI hydrophones, the sensitivity  in the frequency 
range from 10 to 50\,kHz
is higher but less uniform, both as a function of
frequency and as a function of polar angle.

\begin{figure}[tbh]
\centering

\includegraphics[width=12.0cm]{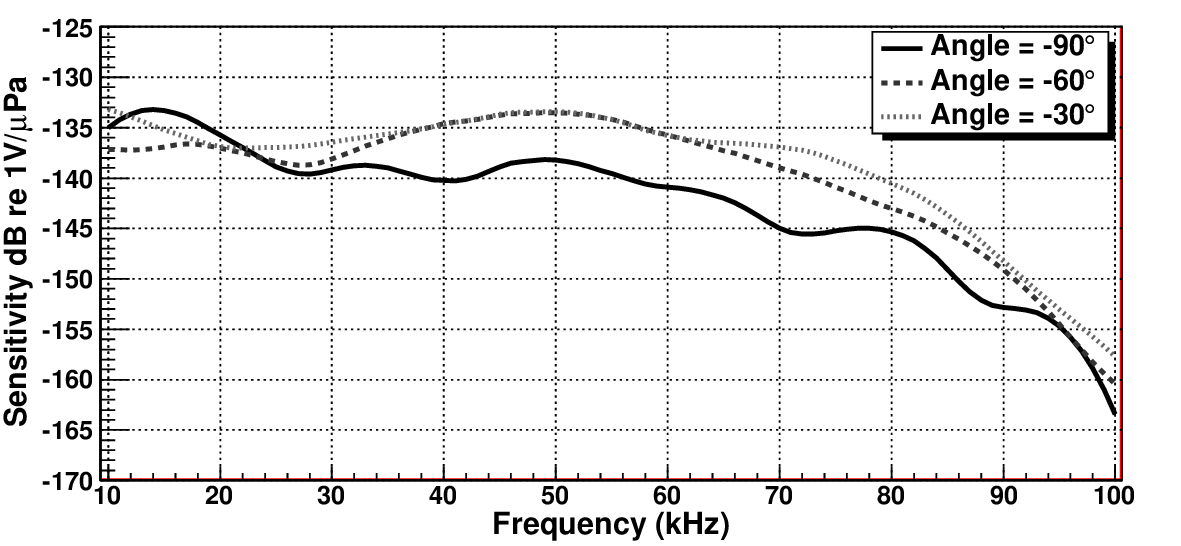}
%\vspace{-1mm}

\includegraphics[width=12.0cm]{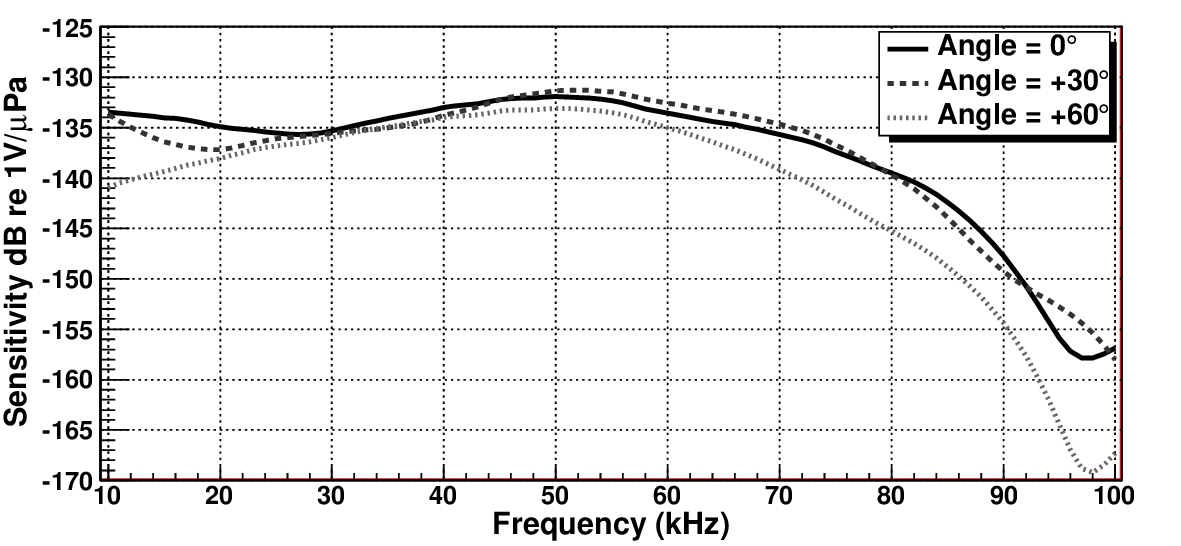}
%\vspace{-1mm}

\caption{
Sensitivity of an AM sensor
as a function of frequency for different angles with respect to the
longitudinal axis of the sensor. 
Systematic uncertainties below 50\,kHz are 2 to 3\,dB. 
}
\label{fig:sensitivity_AM}
\end{figure}

As a consequence of their design, the solid angle over which the sensors of
the AMs are sensitive is smaller compared to the hydrophones.
Furthermore,
reflections and resonances within the glass sphere affect the signal
shape and frequency dependence, making laboratory measurements more
difficult to interpret.  
The calibration was performed by varying the position of the emitter
along a half circle, such that each emitter position has the same
distance to the piezo element.  Angles were then given by the position
of the emitter along the half circle with respect to the longitudinal
axis of the piezo sensor, which defined the angle of 0$^\circ$.
Results are shown in Figure~\ref{fig:sensitivity_AM}. The higher
sensitivity compared to the ECAP hydrophones is due to the different
piezo element that is used and the acoustic coupling between water,
the glass sphere and the piezo sensor.

All sensitivity measurements were done at normal pressure.  A
verification with an in situ calibration has not yet been carried out
at the time of the writing of this paper.

\begin{figure}[ht]
\centering
\includegraphics[width=12.0cm]{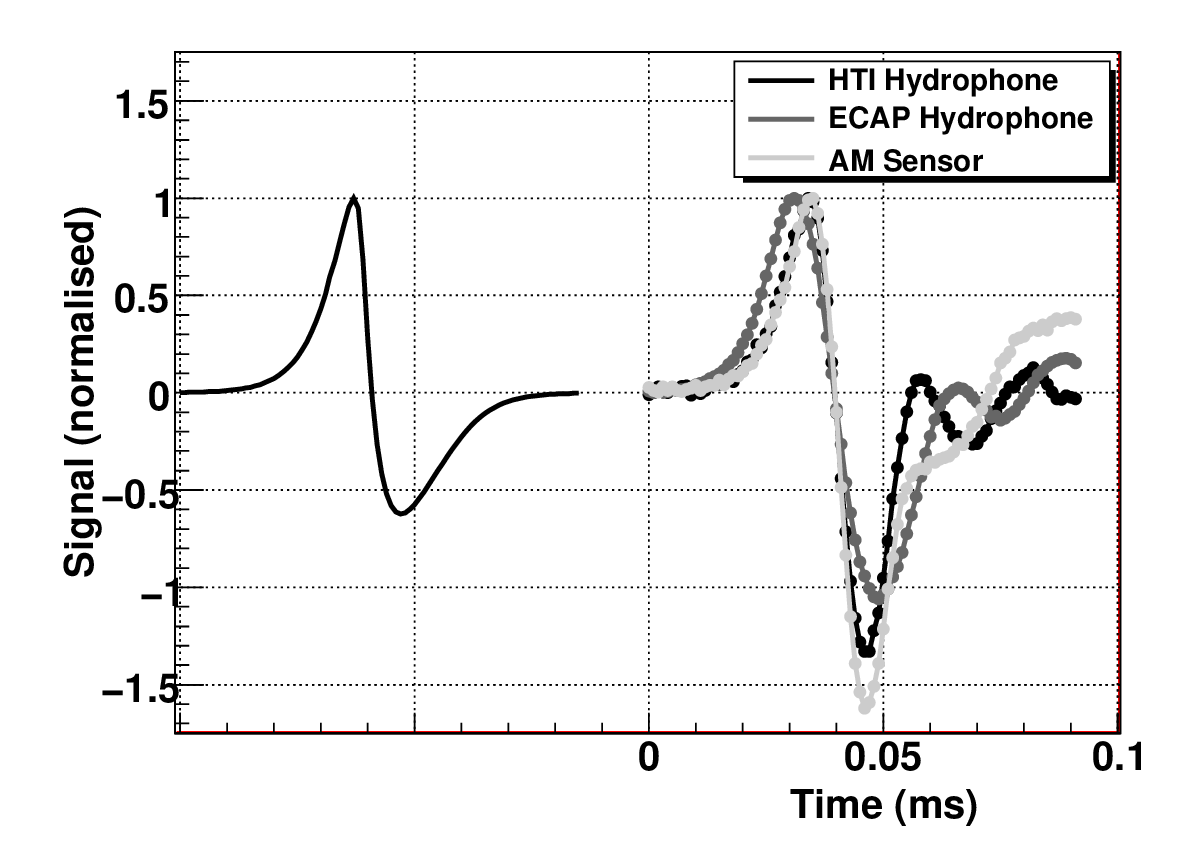}
\caption{Comparison of the response of different acoustic
  sensor types to a bipolar pulse.  The emitted signal is shown on the
  left.  The response of the hydrophones was measured for a polar
  angle of 90$^\circ$, the response of the AM sensor for an angle of
  0$^\circ$.  The first peak of each pulse (including the emitted one)
  was normalised to 1 and the time axis of each received signal was
  adjusted such that the times of the zero crossings coincide.  The
  time offset between emitted and received pulses in the depiction is
  arbitrary.  }
\label{fig:comparison_bips_calibration}
\end{figure}

For the calibration that was describe above, gaussian signals were
emitted which in the frequency domain cover the range of the
calibration.  In addition, the response of the sensors to bipolar
pulses was recorded. This is shown in
Figure~\ref{fig:comparison_bips_calibration}.  The agreement between
the different sensor types is quite good.  The asymmetry of the input
pulse, i.e. the ratio of the pulse heights at the positive and
negative peaks, can be seen to be diminished in the response of the
sensors. This is due to the excitation of oscillations of the piezo
elements.

\subsection{Cables and Connectors}
\label{sec:cables}

Each electronics container is equipped with three 12-pin SubConn
connector sockets\footnote{MacArtney Underwater Technology group,
  http://www.subconn.com/}.
In order to connect two
hydrophones to each of the three sockets, special {\em fanout
  cables} were produced (see Figure~\ref{fig:acou_storey_drawing}).
To the electronics container end of the cable, the same mating 
connector plugs are used as for the Optical Modules. 
At the other end of the cable, a bulkhead connector AWQ-4/24 of the
ALL-WET split series by Seacon\footnote{Seacon (Europe) LTD, Great
  Yarmouth, Norfolk, UK, http://www.seaconeurope.com/} was moulded,
which fans out into six wedge-shaped sectors.  Each sector has a 4-pin
connector socket, serving the four leads for individual power supply
and differential signal readout of the hydrophones. Each of the mating
4-pin connectors is moulded to a neoprene cable with the hydrophone at
its other end. The remaining four sectors of the bulkhead connector
are sealed with blind plugs.

The standard cables used in the ANTARES detector between the
electronics container and the Optical Modules are also used to connect
the AMs. The LCMs integrated into storeys with AMs and with
hydrophones are interchangeable.  All connections and cables within
AMADEUS are functioning as expected.

\subsection{Off-Shore Electronics} 
\label{sec:offshore}

In the ANTARES data acquisition (DAQ) scheme~\cite{bib:antares_daq},
the digitisation is done within the off-shore electronics
container (see Section~\ref{sec:amadeus_overview}).
Each LCM contains a backplane that is equipped with sockets for
the electronics cards and provides them with power and data lines.
A standard LCM for processing the data from PMTs
contains the following electronics boards:
\begin{itemize}
\item
Three {\em ARS motherboards}, each comprising two Analogue Ring Sampler (ARS)
ASICs, for conditioning and digitisation of the analogue signals from
the PMTs~\cite{bib:ARS_paper2};
\item
A {\em DAQ board}, which reads out the ARS motherboards
and handles the communication to shore via TCP/IP;
\item
A {\em Clock board} that provides the timing signals to correlate
measurements performed in different storeys
(see Subsection~\ref{sec:DAQ-and-clock});
\item
A {\em Compass board} to measure the tilt and the heading of
the storey.
\end{itemize}

The transmission of data to shore is done through a {\em Master LCM
  (MLCM)} which---in addition to the components of an LCM described
above---contains an ethernet switch and additional boards for handling
incoming and outgoing fibre-based optical data transmission.  Up to
five storeys form a {\em sector}, in which the individual LCMs
transmit the data to the MLCM.

\begin{figure}[ht]
\centering
\includegraphics[width=12cm]{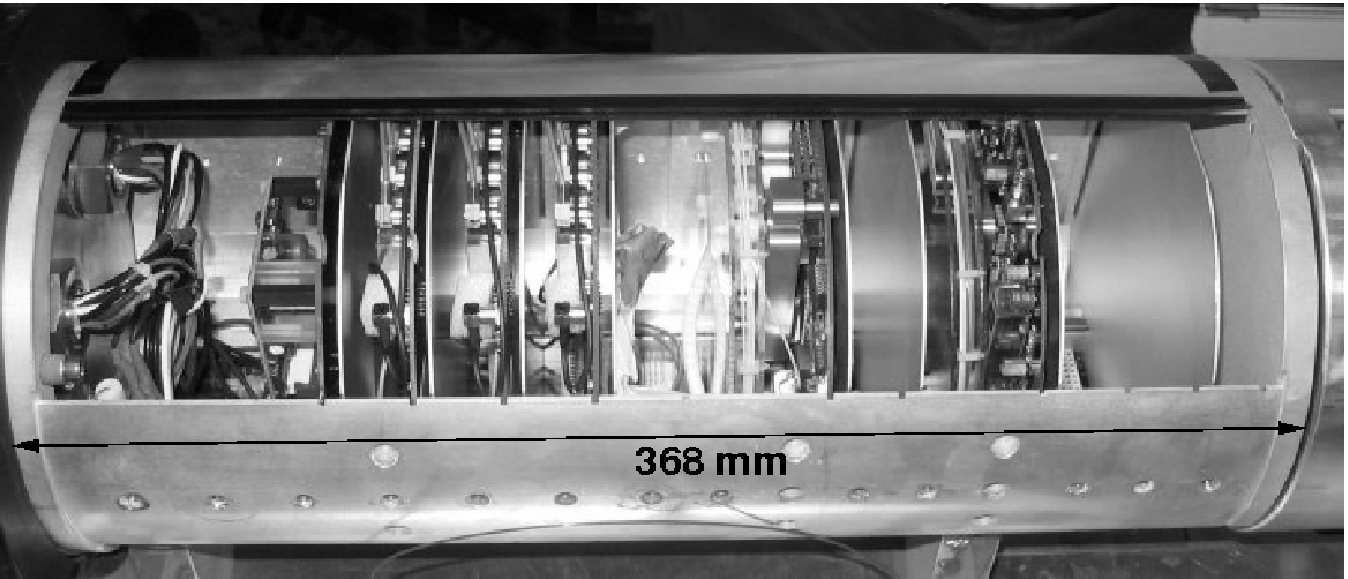}
\hspace{2pc}%
  \caption{ An LCM during assembly, equipped with AcouADC boards
    before insertion into the titanium container.  The sockets for
    external connection (not visible in this picture) are attached to
    the lid of the container on the left-hand side of the photograph.
    From left to right, the following boards are installed: a Compass
    board; three AcouADC boards; a DAQ board; a Clock board.
\label{fig_Acou_LCM}}
\end{figure}

For the digitisation of the acoustic signals and for feeding them into
the ANTARES data stream, the {\em AcouADC board} was
designed.  These boards are pin-compatible with the ARS motherboards
and replace them in the acoustic storeys.
Figure~\ref{fig_Acou_LCM} shows the fully equipped LCM of an acoustic
storey.  

\subsection{The AcouADC Board}
\label{sec:acouadc-board}

Each AcouADC board serves two acoustic sensors and has the following
major tasks:
\begin{itemize}
\item
Pre-processing of the analogue signals (impedance
matching, application of an anti-alias filter, selectable gain
adjustment); 
\item
Digitisation of the analogue signals and preparation of the digitised
data stream for transmission to the DAQ board;
\item
Provision of stable low-noise voltage (6.0\,V) for power supply of the
acoustic sensors;
\item 
Provision of an interface to the on-shore slow control software 
(see Subsection~\ref{sec:SC}).
\end{itemize}

\begin{figure}[ht]
\centering
\includegraphics[width=7.5cm]{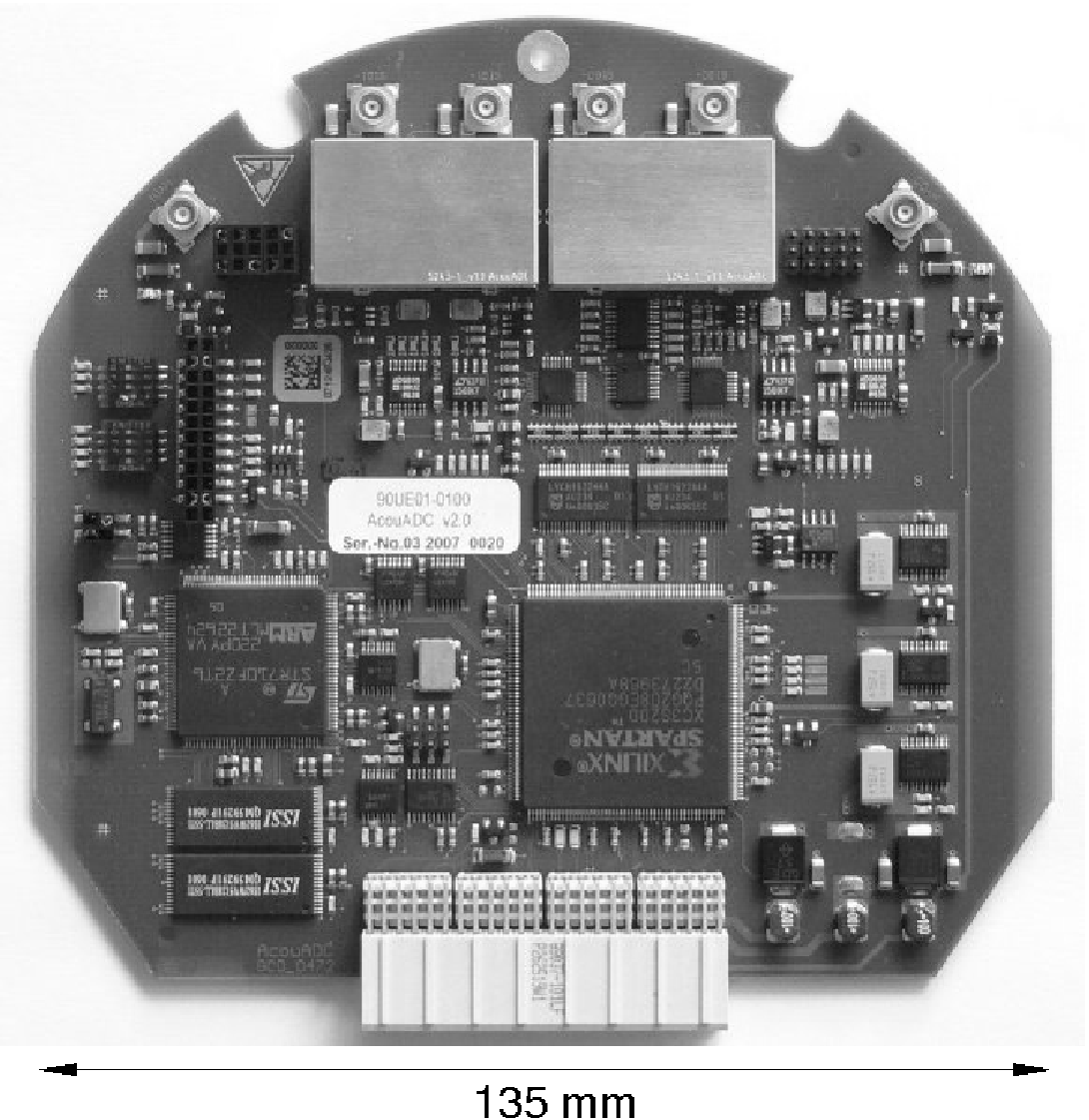}\hspace{2pc}%
  \caption{ Photograph of an AcouADC board. The four connectors for
    the two differential input signals are located at the top, the
    analogue signal processing electronics is covered by metal
    shields.
\label{fig_AcouADC_board}}
\end{figure}

\begin{figure}[tbh]
\centerline{
\includegraphics[width=\textwidth]{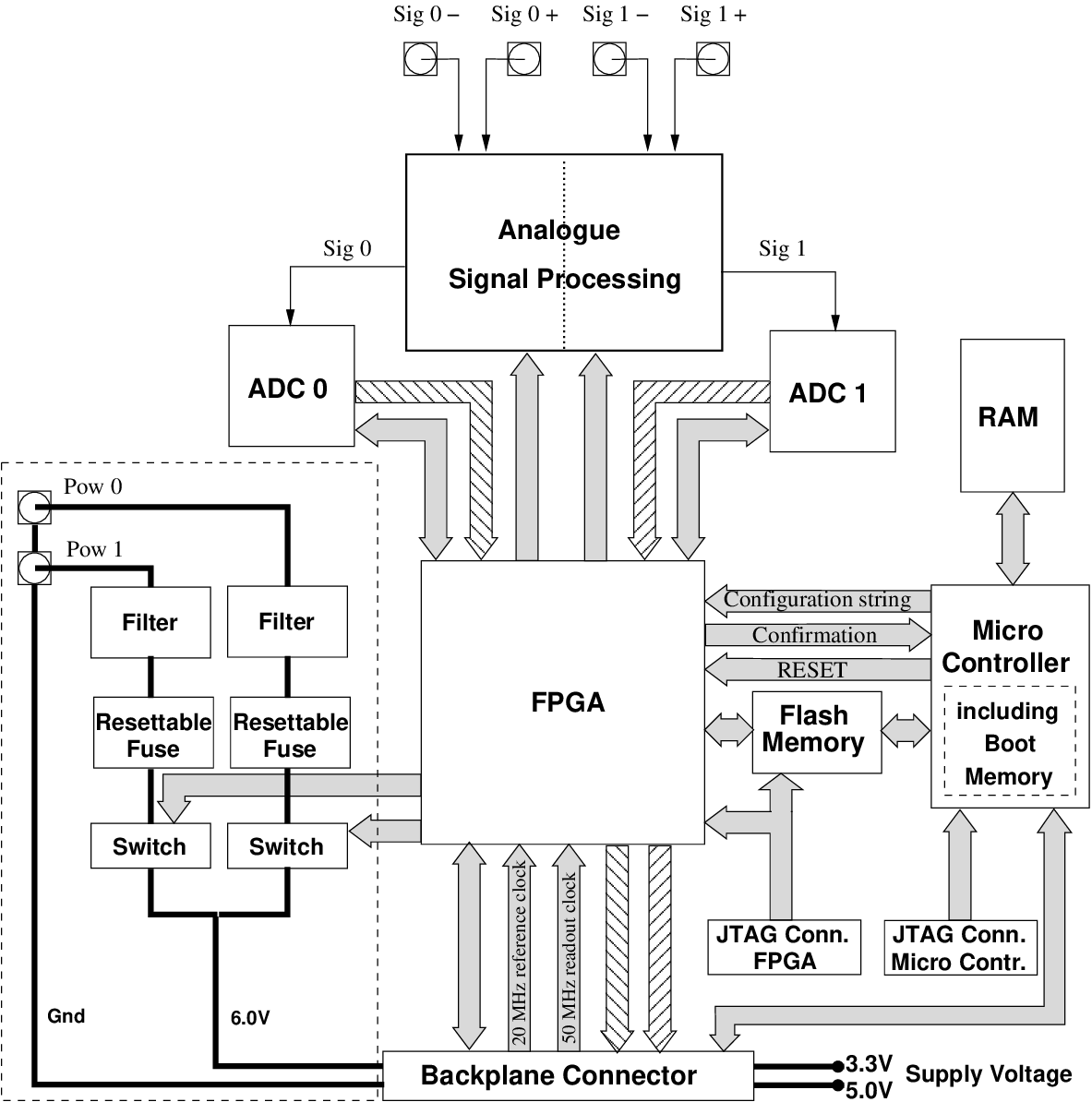}
}
\caption{ Block diagram of the AcouADC board. The flow of the analogue 
  sensor signals entering from the connectors (squares with inserted
  circles) at the top of the figure is indicated by thin arrows.
  Hatched arrows denote the flow of the digitised data further
  downstream.  General communication connections are shown as shaded
  arrows. 
  The components relevant
  for the power supply of the hydrophones are shown in the left part.
  Voltage supply lines are shown as thick lines, the dashed box 
  indicates
  the circuitry for the power supply of the acoustic sensors.
\label{fig-synoptic}
}
\end{figure}

A photograph and a block diagram of an AcouADC board are shown in
Figures~\ref{fig_AcouADC_board} and \ref{fig-synoptic}, respectively.
The board consists of an analogue and a digital 
signal  processing part.  Each board
processes the differential voltage signals from two acoustic sensors,
referred to as ``Sig 0'' and ``Sig 1'' in the diagram.  The two
signals are processed independently and in parallel for the complete
(analogue and digital) data processing chain.

A main design criterion for the board was a low inherent noise level,
so that even for sea state 0 no significant contribution to the
recorded signal originates from the electronics of the board.  To
protect the analogue parts from potential electromagnetic
interference, they are shielded by metal covers.  Tests of the
electromagnetic compatibility (EMC) of the board have shown that this
design is not significantly affected by electromagnetic noise even for
conditions that are far more unfavourable than those present in
situ~\cite{bib:Graf_PhD_2008}.

The two 6.0\,V power supply lines on each AcouADC board (connectors
labelled ``Pow 0'' and ``Pow 1'' in Figure~\ref{fig-synoptic}) are
protected by resettable fuses against short circuits that could be
produced by the sensors due to water ingress. In addition, each voltage
line can be individually switched on or off.

\subsubsection{Analogue Signal Processing}
The analogue signal is amplified in two stages.  The first stage
applies a coarse gain with nominal amplification factors of 1, 10 or
100.  It is implemented as a differential amplifier with single-ended
output, referenced to 2.5\,V.  The gain factor 1 is used for recording
dedicated runs with large signal amplitudes, e.g. from the emitters of
the ANTARES acoustic positioning system
(see Subsection~\ref{sec:amadeus_part_antares}), whereas the factor of
100 was only foreseen for the case that the sensitivity of the
hydrophones would degrade after deployment.

The second amplification stage, the fine gain, 
is intended to adjust the gains of different types of hydrophones. 
It is
a non-inverting amplification with single ended output and a reference
voltage of 2.5\,V.  Gain factors of 1.00, 1.78, 3.16, and 5.62 
(corresponding to 0, 5, 10, and 15\,dB, respectively) are selectable by 
switching between four appropriate resistors in the feedback loop 
of the operational amplifier. 
Combining the two stages, the gain factor can be set to one of 12 values
between 1 and 562.  The standard setting is an overall gain factor of
10, yielding the approximate sensitivity of $-$125\,dB\,re\,1V/$\upmu$Pa. 

After amplification, the signal is coupled into a linear-phase
10th-order anti-alias filter\footnote{Filter LTC1569-7 from Linear
  Technology, http://www.linear.com/} with a root-raised cosine
amplitude response and a 3\,dB point at a frequency of 128\,kHz.  In
low-power mode, the filter output range is about 3.9\,V.  The output
is referenced to 2.0\,V and fed into a 16-bit analogue-to-digital
converter (ADC) that will be described below.
Accordingly, the ADC reference voltage is set to 2.0\,V, implying
that the digital output of zero corresponds to this analogue value.
The input range of the ADC is 0.0 to 4.0\,V.

The three analogue stages (coarse and fine amplification and
anti-alias filtering) and the ADC are decoupled by appropriate
capacitors.
Furthermore, several RCL elements within the analogue signal chain form
an additional band-pass filter. Its low-frequency 3\,dB point is at
about 3\,kHz and cuts into the trailing edge of the low-frequency noise of
the deep-sea acoustic background~\cite{urick}, protecting the
system from saturation.
The high-frequency 3\,dB point is above 1\,MHz and was introduced to
comply with the input requirements of active components of the
circuitry.

\subsubsection{Digital Signal Processing}
The digital part of the AcouADC board digitises the acoustic signals
and processes the digitised data.
%%%%
It is flexible due to the use of a micro controller
($\upmu$C)\footnote{STR710 from STMicroelectronics,
  http://www.st.com/} and a field programmable gate array
(FPGA)\footnote{Spartan-3 XC3S200 from XilinX, http://www.xilinx.com/}
as data processor.  All communication with the shore is done via the
DAQ board; the $\upmu$C handles the slow control (see Subsection
\ref{sec:SC}) and the FPGA the data transfer.  The $\upmu$C is
accessed from the on-shore control software and is used to adjust
settings of the analogue part and the data processing.  It can also be
used to update the firmware of the FPGA.  This firmware is stored in a
flash memory and loaded after a reset of the FPGA.  In situ, this reset is
asserted from the $\upmu$C.  If a firmware update is performed, the
$\upmu$C first loads the code from the shore into the random access
memory (RAM). Once the integrity of the code has been confirmed by
means of a checksum, the code is transmitted into the flash memory. In
order to avoid the potential risk that a software error renders the
$\upmu$C inaccessible, its boot memory can only be changed in the
laboratory.  For testing and programming in the laboratory,
JTAG\footnote{ The IEEE standard 1149.1 ``Test Access Port and
  Boundary-Scan Architecture'' is commonly referred to as JTAG, the
  acronym for Joint Test Action Group. It defines an interface to
  individual components on a circuit board.}  is used to access the
FPGA, the $\upmu$C, and the flash memory.

For each of the two input channels, the digitisation is done at
500\,kSps (kilosamples per second) by one 16-bit successive
approximation ADC\footnote{ ADS8323 from Analog Devices,
  http://www.analog.com/} with output range from $-$32768 to $+$32767
counts.  The two ADCs are read out in parallel by the FPGA and further
formatted for transmission to the DAQ board.

ADCs commonly show relatively high deviations from a linear
behaviour near the zero value of their digital output.  The size of
this effect depends on the circuitry into which the ADC is embedded.
For the prototypes of the AcouADC boards, this effect proved to be
fairly pronounced. For this reason, the reference voltage of the
anti-alias filter output can be switched from its standard value of
2.0\,V to 1.0\,V, thereby moving the centre  of the acoustic noise
distribution away from the digital value of zero.
This leads to an asymmetry of the recordable range for
positive and negative amplitudes of acoustic signals, effectively reducing
the dynamic range by a factor of 2 if one requires both positive and negative 
amplitudes to be fully recorded.
For standard data taking, this is not desirable 
whereas the non-linearities of the ADCs proved to be unproblematic.
Therefore, the reference voltage of the
anti-alias filter output is set to  1.0\,V only for special measurements.

In standard mode, the sampling rate is reduced to 250\,kSps in the
FPGA, corresponding to a downsampling by a factor of 2 (DS2).
Currently implemented options are DS1 (i.e. no downsampling), DS2, and
DS4, which can be selected from the shore.  For each downsampling
factor an adapted digital anti-alias filter, compliant with the
Nyquist-Shannon sampling theorem, is implemented in the FPGA as a
finite impulse response (FIR) filter with a length of 128 data points.
For DS2, the frequency spectrum between the 3\,dB points at 2.8 and
108.8\,kHz passes the filter.

\subsubsection{System Characteristics}
\begin{figure}[ht]
\centering
\includegraphics[width=10.0cm]{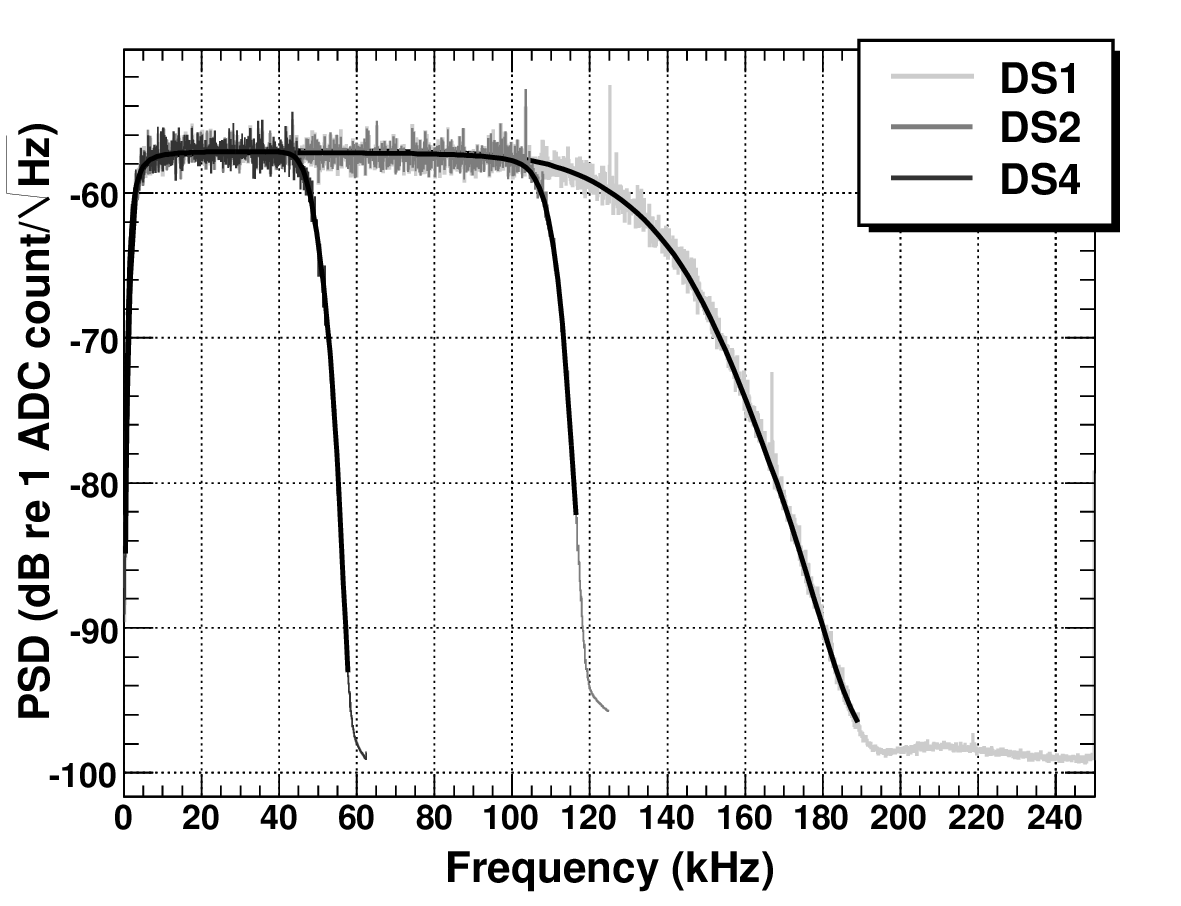}
\caption{The AcouADC board filter response, characterised by a power
  spectral density (PSD) as a function of frequency, measured for the
  three different downsampling factors. For each of the three
  measurements, the parametrisation is shown as a black line.
\label{fig:freqresp_amp}
}
\end{figure}

The complex response function of the AcouADC board (i.e. amplitude and
phase) was measured in the laboratory prior to deployment for each
board and a parametrisation of the function was
derived~\cite{bib:Graf_PhD_2008}.
Figure~\ref{fig:freqresp_amp} shows the frequency response of the
AcouADC board.  The measurement was done by feeding gaussian white
noise into the system and analysing the digital output recorded by the
board.  Without downsampling (DS1), the rolloff at high frequencies is
governed by the analogue anti-alias filter. For DS2 and DS4, the
digital FIR filters are responsible for the behaviour at high
frequencies.  At low frequencies, the effect of the band-pass filter
described above can be seen.
Figure~\ref{fig:freqresp_amp} furthermore demonstrates that within each
passband, the filter response is essentially flat.  The comparison of
the recorded data with the parametrisation shows excellent agreement.

\begin{figure}[ht]
\centering
\includegraphics[width=10cm]{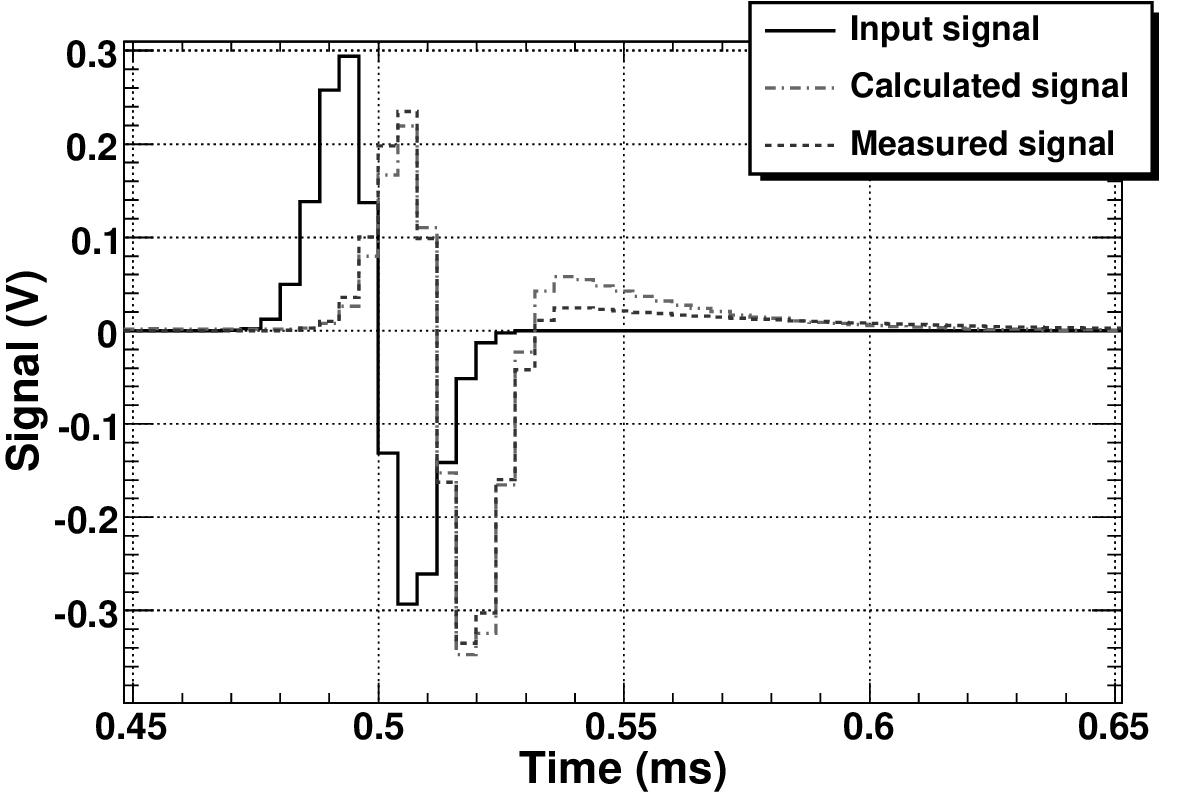}
\caption{The response of the analogue part of an AcouADC board to a
  bipolar input pulse. Shown are the measured output signal and the
  output signal calculated from the parametrised response
  function. The measured output signal was obtained with an
  oscilloscope at the input of the ADC.  The measurement was done for
  a nominal gain factor of 1.}
\label{fig:signal_response_acouADC}
\end{figure}

The parametrisation of the response function allows to calculate the
response of the system to any input pulse and vice versa the
reconstruction of the original shape of any recorded pulse.
Figure~\ref{fig:signal_response_acouADC} shows a comparison of the
measured and calculated response of the analogue signal processing
part of the AcouADC board to a 
generic bipolar input pulse
as it would be expected from a
neutrino-induced cascade 
(see Section~\ref{sec:introduction}).
The digital FIR filter introduces an additional time offset of 
128\,$\upmu$s of the digitised data for DS2 and DS4.

The ADCs of the AcouADC board were investigated in
detail~\cite{bib:Graf_PhD_2008}.  For each individual ADC, the
transfer curve from input voltage to ADC counts was measured and
distortions from the ideal linear behaviour quantified in terms of the
differential nonlinearity (DNL) and integral nonlinearity (INL).  The
distribution of the DNL values for all bins of all calibrated signal
channels shows negligible deviations from the ideal situation (i.e. a
peak at zero) with a mean of $-$0.02 ADC counts and a standard
deviation of 0.06 ADC counts, corresponding to 3.4\,$\upmu$V. The
values of the INL of the ADCs stay within ±50 ADC counts for all
signal channels over the full input range, corresponding to
$\pm$3.1\,mV.

The spurious-free dynamic range (SFDR) of an ADC is a measure of its
dynamic range.  Using a sinusoidal input signal, the average SFDR of
the ADCs of all boards in AMADEUS was measured to be 59.9$\pm$1.1\,dB,
meaning that harmonics of the sine wave distorting the output signal
are suppressed by 3 orders of magnitude in the amplitude.  Hence, a
clear determination of the frequency is possible even for saturated
signals, for which the harmonic components are enhanced.

The 12 gain factors for each channel were calibrated and the
correction factor for gain 1 was measured to be 1.05$\pm$0.01.  The
gain factors of the other 11 settings were measured with respect to
this value and were found to deviate from the nominal factors by
about 10\% at maximum.

The inherent noise of the electronics (output for open signal input)
and the cross-talk between the two signal channels of an AcouADC board
were measured to be negligible in
comparison with the inherent noise of the acoustic sensors.

\subsection{Slow Control System}
\label{sec:SC}

The ANTARES slow control (SC) system has two main tasks. It provides
the off-shore components with initialisation and configuration
parameters and it regularly monitors whether the operational
parameters are within their specified ranges.  In addition, the readout
of some instruments for environmental
monitoring~\cite{bib:milom-paper}, performed at intervals of a few
minutes, is polled and sent through the SC interface.

For the AMADEUS system, the following parameters can be set from shore
via the SC system for each acoustic channel individually: one of the
12 values for the gain; downsampling factors of 1, 2, or 4 (or no data
transmission from the AcouADC board); power supply for the acoustic
sensor on or off; reference voltage of the analogue signal fed into
the ADC 2.0\,V or 1.0\,V.

To monitor the environment within each LCM container, a humidity
sensor and temperature sensors on several boards are installed. One
temperature sensor is placed on each AcouADC board.
Values read out by the SC system are stored in an
Oracle\raisebox{.6ex}{\scriptsize \textregistered} database that is
centrally used by ANTARES.

\subsection{Data Acquisition and Clock System}
\label{sec:DAQ-and-clock}
AMADEUS uses the same DAQ system and follows the same ``all data to
shore'' strategy~\cite{bib:antares_daq} as the ANTARES neutrino
telescope, i.e. all digitised data are transmitted to shore via
optical fibres using the TCP/IP protocol.  The data stream from the
sender DAQ board is tagged with the IP address of the receiving
on-shore server. In the control room, the acoustic data are routed to
a dedicated computer cluster by using the transmitted IP address.
The ANTARES clock system operates separately from the DAQ system,
using a different set of optical fibres to synchronise data from
different storeys. The system provides a highly stable 20\,MHz
synchronisation signal, corresponding to a resolution of
50\,ns\footnote{ The much higher precision that is required for the
  synchronisation of the optical signals from the PMTs is provided by
  a 256-fold subdivision of the 20\,MHz signal in the ARS motherboards.},
which is generated by a custom-designed system at the ANTARES control room.
The synchronisation of this internal clock with the
UTC\footnote{Coordinated Universal Time} of the
GPS system is established with a precision of 100\,ns.

The synchronisation signal is broadcast to the off-shore clock
boards and from there transmitted further to the FPGA of the AcouADC
board.  Based on this signal, the data packages sent from the AcouADC
board to shore via the DAQ board receive a timestamp which allows for
correlating the data from different storeys.  The 50\,ns resolution of
the timestamp by far exceeds the requirements given by the standard
sampling time of 4\,$\upmu$s corresponding to DS2.  Differences in the
signal transit times between the shore station and the individual
storeys are also smaller than 4\,$\upmu$s and do not need to be
corrected for.

\subsection{On-Shore Data Processing and System Operation}
\label{sec:onshore}

The AMADEUS system is operated with its own instance of the
standard ANTARES control software called {\em
  RunControl}~\cite{bib:antares_daq}.  This is a program with a
graphical user interface to control and operate the experiment. It is
Java\raisebox{.4ex}{\scriptsize \texttrademark}-based and reads the
configuration of the individual hard- and software components from the
ANTARES database, allowing for an easy adaption of individual run
parameters for the AMADEUS system.  Via the clock system the absolute
time of the run start is logged in the database with the
aforementioned precision of 100\,ns.  The end of a run is reached if
either the data volume or the duration exceed predefined limits (in
which case a new run is started automatically) or the run is stopped
by the operator.  The data of one AMADEUS run are stored in a single
file in {\em ROOT} format~\cite{bib:root}. The typical duration of a
run ranges from 2 to 5 hours.

For the computing requirements of AMADEUS, a dedicated on-shore
computer cluster was installed.  It currently consists of four
server-class computers, of which two are used for data
triggering\footnote{While this functionality might be more commonly
  referred to as filtering, it is ANTARES convention to refer to the
  ``on-shore trigger''.}  (equipped with 2$\times$dual core 3\,GHz
Intel Xeon 5160 and 2$\times$quad core 3\,GHz Intel Xeon 5450
processors, respectively).
Hence, a total of 12 cores are available to process the data.
One of the remaining two computers is used to write the data to an
internal 550 GB disk array (RAID), while the other is used to operate the
RunControl software and other miscellaneous processes and to provide
remote access to the system via the Internet.

The AMADEUS trigger searches the data by an adjustable software filter;
the events thus selected are stored to disk. This way the raw data
rate of about 1.5\,TB/day is reduced to about 10\,GB/day for storage.
Currently, three trigger schemes are in operation~\cite{Neff_diplom}:
A minimum bias trigger which records data continuously for about 10\,s
every 60\,min, a threshold trigger which is activated when the signal
exceeds a predefined amplitude, and a pulse shape recognition
trigger. For the latter, a cross-correlation of the signal with a
predefined bipolar signal, as expected for a neutrino-induced cascade,
is performed. The trigger condition is met if the output of the
cross-correlation operation exceeds a predefined threshold.  With
respect to a matched filter, this implementation reduces the run time
complexity while yielding a comparable trigger performance.

As discussed in Section~\ref{sec:introduction}, for pressure pulses
induced by neutrino interactions the amplitude, asymmetry and
frequency spectrum depend on the position of the observer with respect
to the particle cascade.
The predefined bipolar signal used for the pulse shape recognition
trigger corresponds to the pulse shape expected at a distance of roughly
300\,m from the shower centre in the direction perpendicular to the
shower axis, i.e. where the maximum signal within the flat volume of
sound propagation is expected.
The cross correlation with pulses whose shape differs from the implemented
one changes the peak in the cross correlation output: it is broadened and
diminished as compared to the filter response on the predefined signal. This
effectively increases the trigger threshold in terms of
pressure amplitude for such pulses.
As will be described below, the final trigger decision requires coincidences
within an acoustic storey, which allows the 
trigger threshold for the cross correlation output of each individual 
acoustic sensor to be set to a low value.
Given that the main purpose of the AMADEUS system is the investigation of 
background  noise, this implementation is very efficient in
recording a wide range of bipolar and multipolar events.
Dedicated 
searches for neutrino signals---which are difficult due to the geometry
of the acoustic storeys within the AMADEUS system---are done
offline, taking into account the variations of the pulse shapes with
distance and direction.

Both the threshold and the pulse shape recognition trigger are
applied to the individual sensors and are self-adjusting to the
ambient noise, implying that all trigger thresholds are defined in
terms of a signal-to-noise ratio. 
The trigger thresholds are software parameters and therefore can be set at
will.
The noise level is calculated from 
and applied to the data of the {\em
frame} that is currently being analysed. A frame denotes the structure
in which data are buffered off-shore by the DAQ board before being
sent to shore and contains data sampled during an interval of about
105\,ms~\cite{bib:antares_daq}.  
If one of these two trigger conditions is met, an additional trigger
condition is imposed, which requires coincidences of a predefined
number of acoustic sensors on each storey.  The coincidence window is
fixed to the length of a frame.  Currently, the coincidence trigger
requires that the threshold or pulse shape recognition trigger
conditions have been met for at least four out of six sensors of a
storey.

In the ANTARES DAQ system, the frames start at fixed intervals with
respect to the run start.  Trigger conditions are imposed on
temporally corresponding frames from all storeys simultaneously,
whereupon the frames are discarded and data not selected by the
trigger are lost. Processing subsequent frames at the same time is not
possible.  Given the distances of typically 1\,m between sensors
within one storey, time delays between signals from a given source are
always less than 1\,ms.  Therefore, the number of sources for which
the signals extend over two frames, and hence the coincidence trigger
may not be activated, is small.  The disadvantage of a large trigger
window, the increased probability for random coincidences, leads only
to a small increase of the recorded data volume.
The coincidence trigger can be optionally extended to require
coincidences between different storeys on the same line.  With
distances between storeys ranging from about 10 to 100\,m
(corresponding to delays of the order of 10 to 100\,ms) signals
originating from above or below are suppressed.  This trigger is
currently not enabled.

The data of all sensors  that have fired a coincidence trigger are stored
within a common time window that covers all triggered signals.
Its minimum length is 1.536\,ms for 4\,$\upmu$s sampling time 
(corresponding to 384 data samples) and its maximum length corresponds 
to the length of a frame.

The triggers of the AMADEUS system and the main ANTARES optical
neutrino telescope are working completely independently. Hence, the
search for correlated signals relies on offline analyses.

Both the off-shore and on-shore part of the  
AMADEUS system are scalable, rendering it very flexible.
The enlargement of the system from three to six acoustic storeys with
the commissioning of Line 12
(see Subsection~\ref{sec:amadeus_part_antares}) was easily implemented by
increasing the on-shore computing power and updating the control
software.  In principle, the system could be upgraded to much bigger
numbers of acoustic storeys.  To reduce the data volume transmitted to
shore, it would also be possible to move parts of the trigger
algorithms into the FPGA of the AcouADC board.

AMADEUS is controlled remotely via the Internet. 
Data are centrally stored and are remotely available.

\section{System Performance}
\label{sec:sysperformance}

\subsection{General}

AMADEUS is continuously operating and taking data with only a few
operator interventions per week. The
up-time of each sensor is typically better than 80\%. 
Its ability to continuously send unfiltered data, sampled at high
frequency, to shore for further analysis renders the AMADEUS system a
multipurpose apparatus for neutrino feasibility studies, acoustic
positioning and marine research.

The concept of acoustic clusters (i.e. the acoustic storeys) is very
beneficial for fast online processing.  By requiring coincident
signals from at least four sensors within a storey, the trigger rate
is significantly reduced, improving the purity of the sample selected
with the pulse shape recognition trigger.

The parallel operation of two separate RunControl programs for AMADEUS
and the main ANTARES neutrino telescope, which was originally not
foreseen, has proven to be very successful.  No interference between
the two programs has been observed while the two systems can optimise
their detection efficiency and respond to potential problems almost
independently. At the same time, both systems profit in the same
fashion from developments and improvements of the RunControl software
and monitoring tools.

The stability of the system response is excellent. This was verified
prior to deployment as well as in situ. It was quantified by observing
the mean of the ambient noise distribution as a function of
time. In situ, the 10\,s of continuous data recorded every hour with
the minimum bias trigger were used to calculate the mean.  The 
standard deviation of this value for the first year of operation
is less than 2 $\times$ 10$^{-5}$ of the full range. 

All sensor types described in Subsection~\ref{sec:acou_storey} are
well suited for the investigation of acoustic particle detection
methods.  A comparison of the characteristics of the different sensor
types will be drawn below.

\subsection{Ambient Noise}
\label{sec:ambient_noise}

Studies of the power spectral density of the ambient noise
at the ANTARES site have been performed using the
minimum bias trigger data. 
\begin{figure}[tbh]
\centering
\includegraphics[width=10cm]{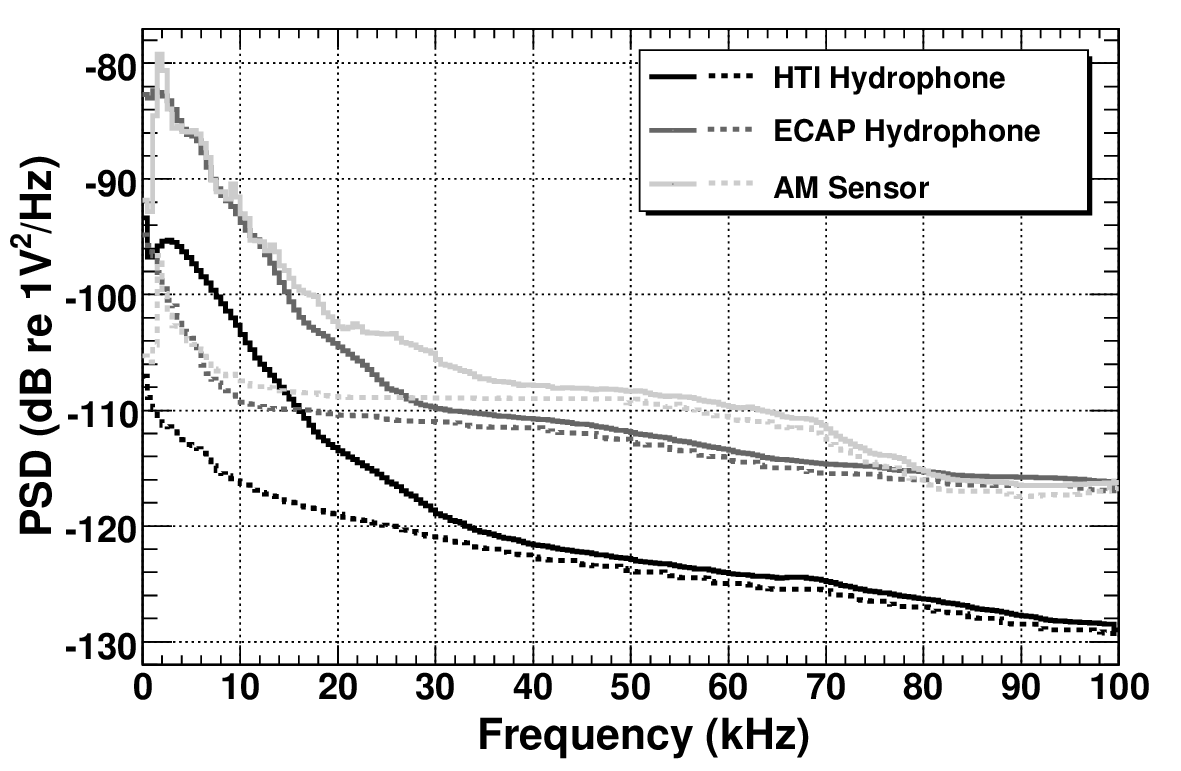}
\caption{ Power spectral density (PSD) of the noise level recorded in
  situ.  Solid lines represent the median, dotted lines the 0.27\%
  quantile (corresponding to a 3$\sigma$ deviation from the median).
  The voltage used is the calibrated input voltage of the AcouADC
  board.
}
\label{fig:sensor_types_sea_noise}
\end{figure}
Figure~\ref{fig:sensor_types_sea_noise} shows the noise levels
recorded with representative HTI and ECAP hydrophones and with one AM
sensor on Line 12 over several time periods with a combined duration
of six months.  For each sensor, the power spectral density was
calculated for the 10\,s intervals of continuous data and then for
each 1\,kHz bin, the median and the 0.27\% quantile (corresponding to
a 3$\sigma$ deviation from the median) were derived for the complete
set of measurements.  For each sensor, one can observe a
characteristic frequency above which the 0.27\% quantile of the noise
level shows a constant difference to the median.  The corresponding
frequencies are about 35\,kHz, 30\,kHz, and 40\,kHz for the HTI
hydrophones, ECAP hydrophones and AM sensors, respectively.  For
higher frequencies, the noise is dominated by the intrinsic
electronics noise, limiting the capability to study the acoustic
noise.

The noise floor is the lowest for the HTI hydrophones, the difference
in the power spectral density to the ECAP hydrophones and AM sensors
being 10 to 15\,dB.  When recording transient signals, the effect on
the signal-to-noise ratio is partially compensated by the higher
sensitivity of the ECAP hydrophones and the AM sensors.
The noise spectrum of the AM sensor displays some structure for
frequencies up to about 25\,kHz.  This is due to coupling of the
sensor to the glass spheres.
In summary, for studies of the in-situ ambient noise, the HTI
hydrophones are the most suited type of sensors.

\begin{figure}[ht]
\centering
\includegraphics[width=10cm]{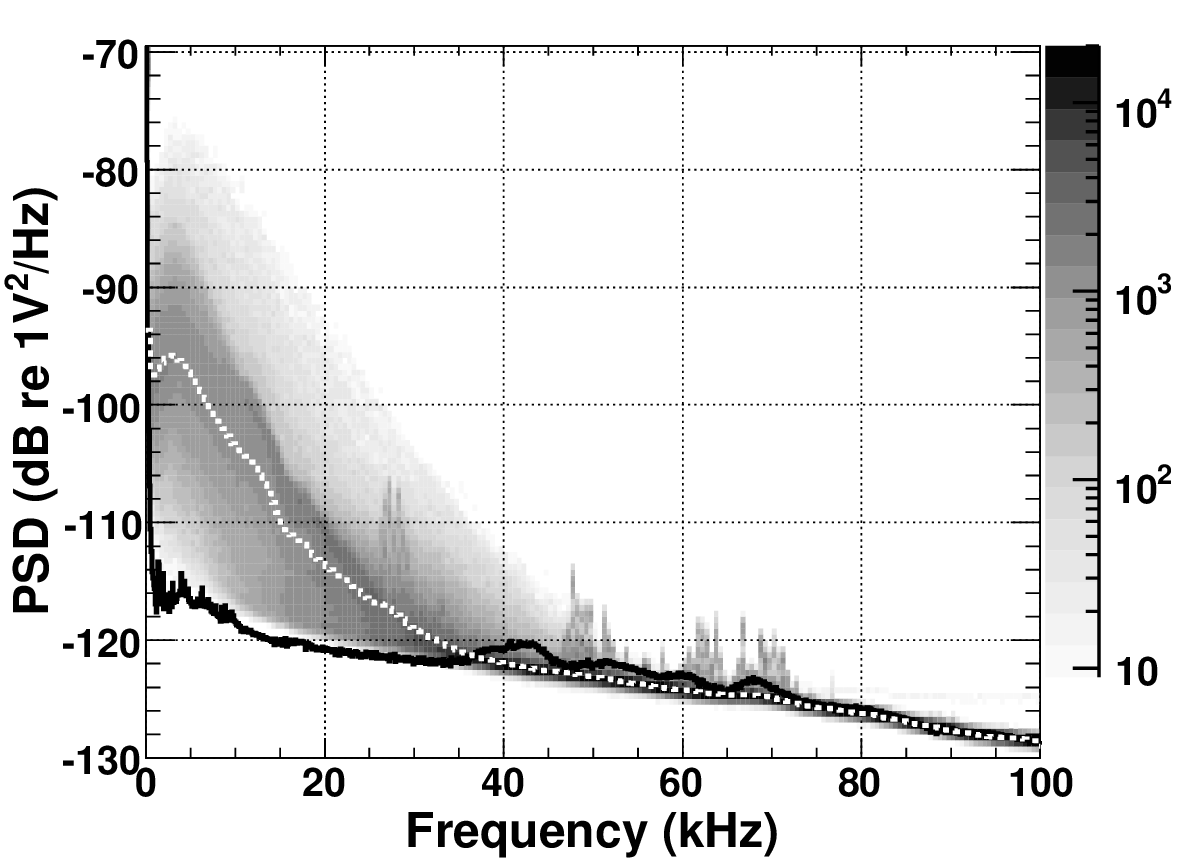}
\caption{Power spectral density (PSD) of the ambient noise recorded
  with one HTI sensor on the topmost storey of the IL.  The voltage
  used is the calibrated input voltage of the AcouADC board.  Shown in
  shades of grey is the occurrence rate in arbitrary units, where dark
  colours indicate higher rates.  Shown as a white dotted line is the
  median value of the in-situ PSD and as a black solid line the noise
  level recorded in the laboratory prior to deployment.  }
\label{fig:ambient_noise}
\end{figure}

In Figure~\ref{fig:ambient_noise} a more detailed presentation of the
noise data recorded with an HTI hydrophone during the year of 2008 is
given.  An algorithm to remove strong transient signals (mostly coming
from the emitters of the acoustic positioning system) was applied. The
relics of such signals and electronics noise show up as spikes between
45 and 75\,kHz.
The lowest level of recorded noise in situ was confirmed to be
consistent with the inherent noise of the system recorded in the
laboratory prior to deployment.  The observed in-situ noise can be
seen to go below the noise level measured in the laboratory for
frequencies exceeding 35\,kHz. This is due to electronic noise
coupling into the system in the laboratory that is absent in the deep
sea.

The overall noise levels (i.e. the RMS of the signal amplitudes in
each 10\,s sample) recorded at the same time with any two active
sensors of the same type are correlated at a level above 90\%.  This
shows that the recorded data are indeed representative of the ambient
conditions and not determined by the inherent noise of the system.

\subsection{Transient Signals and Dynamic Range}
\begin{figure}[ht]
\centering
\includegraphics[width=10cm]{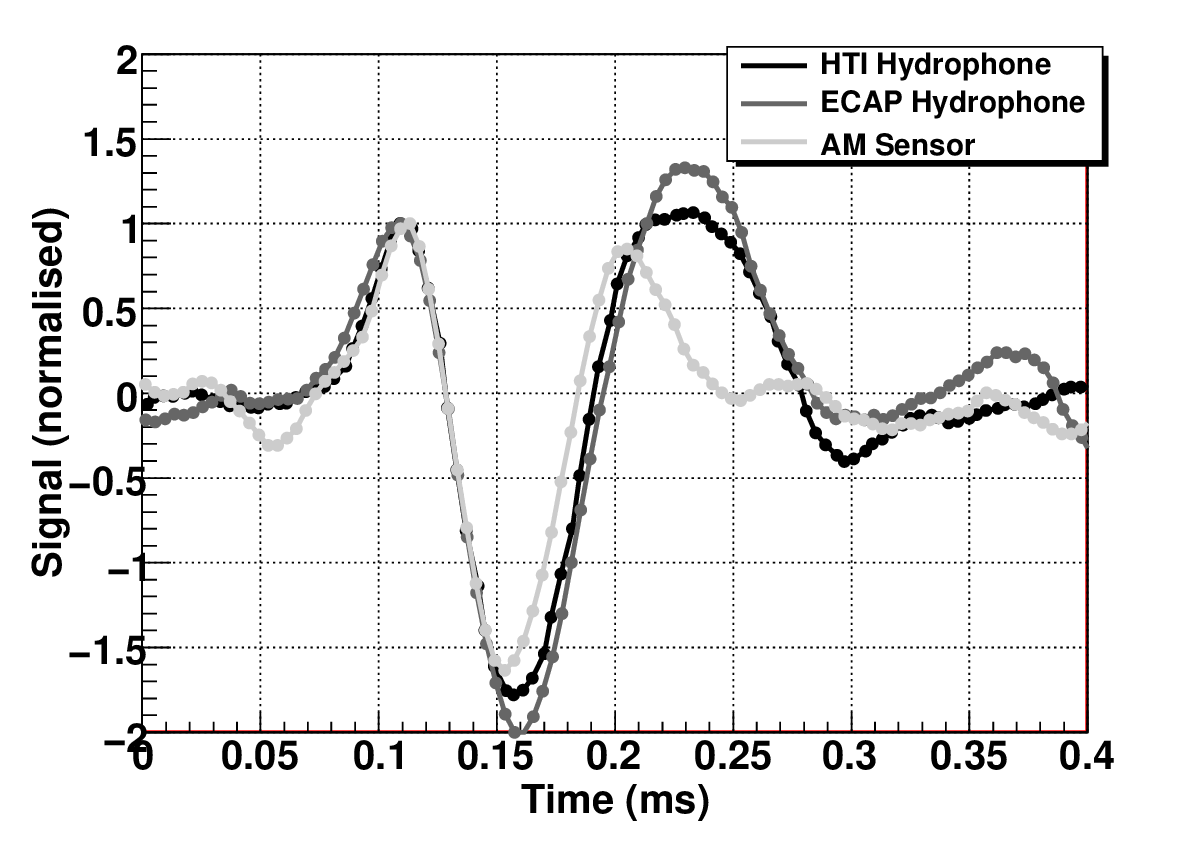}
\caption{ Comparison of the signals originating from a common source
  as recorded by three different types of acoustic sensors, each one
  located on a different storey of Line 12. For better comparability,
  the first peak of each signal has been normalised to 1 and the time
  axis of each signal adjusted such that the times of the zero
  crossings between the first positive and negative peak coincide.  }
\label{fig:bip_comparison}
\end{figure}

The signals recorded with the three different types of acoustic
sensors on Line 12 for a common source is shown in
Figure~\ref{fig:bip_comparison}. The signals were recorded in May 2010
and were received under an angle of about 65$^\circ$ with respect to
the direction pointing vertically upwards.  The agreement between the
signal shapes can be seen to be very good. For the second positive
peak at about 0.20 to 0.25\,s, the AM sensor shows a differing
behaviour from the hydrophones, which can be attributed to the
coupling of the sensor to the glass sphere.

Bipolar signals selected with the pulse shape recognition trigger
typically have a signal-to-noise ratio exceeding 2 for a single
sensor.  Assuming a noise level of 10\,mPa in the frequency range from
1 to 50\,kHz, which is a typical level recorded at calm sea,
 i.e. sea state 0,  a signal of 20\,mPa can be
detected. Such a signal is expected to be emitted from a 2\,EeV
cascade emerging from a neutrino interaction at a distance of about
200\,m~\cite{bib:Sim_Acorne}.
Improving the trigger techniques may further enhance the energy
sensitivity. 
Furthermore, the optimal frequency range must be determined to maximise the
signal-to-noise ratio for pulses stemming from neutrino interactions.
To conclude on the feasibility of a large-scale acoustic neutrino
detector in the deep sea---the main objective pursued with the AMADEUS
system---further detailed studies are required.

The maximal pressure amplitude that can be recorded for a gain factor
of 10 without saturating the input range of the ADC is about 5\,Pa.
Usually, only anthropogenic signals originating close to the detector
reach this pressure level at the positions of the hydrophones.

The direction and position reconstruction of acoustic point sources are
currently being pursued as one of the major prerequisites to identify
neutrino-like signals.  
First results are presented in \cite{bib:Richardt_reco, bib:richardt_ARENA08}.

\subsection{Position Calibration of Acoustic Storeys}

\begin{figure}[ht]
\centering
\includegraphics[width=10.0cm]{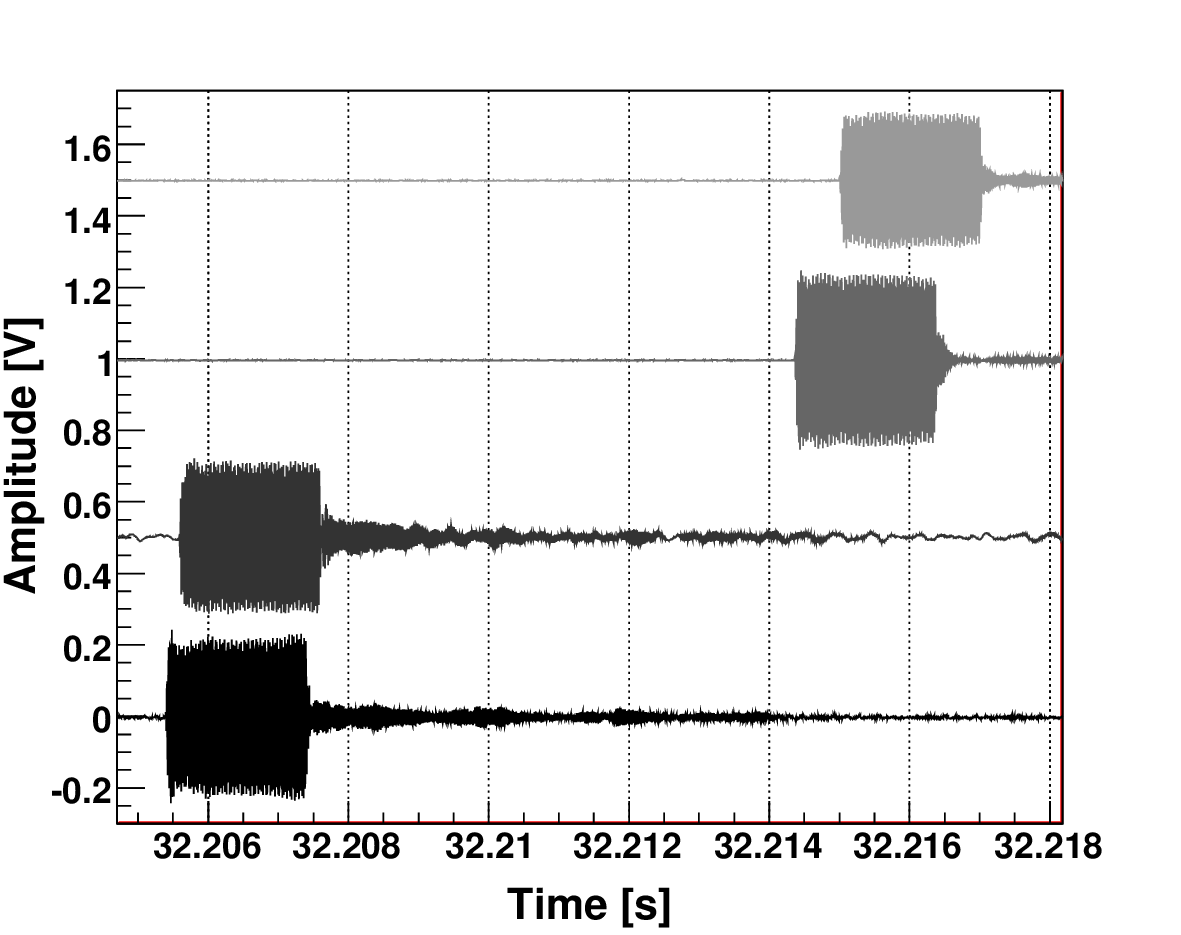}
\caption{ Typical signals of the ANTARES acoustic positioning system
  as recorded with four sensors of the AMADEUS system for gain factor
  1 of the AcouADC board.  For better clarity an offset, starting with
  0\,V and incremented by 0.5\,V with respect to the previous one, is
  added to the amplitude of each sensor.  The first two signals along
  the time axis were recorded by the acoustic storey holding AMs (see
  Figure~\protect\ref{fig:ANTARES_schematic_all_storeys}).  The
  following two signals were recorded with two hydrophones on the
  acoustic storey just above---one hydrophone mounted at the bottom
  and the other one at the top of the storey.  The time is counted
  since the start of the run.
\label{fig:pinger_am_hti}
}
\end{figure}

Just as for the PMTs in the standard storeys, the relative positions
of the acoustic sensors within the detector have to be continuously
monitored.  This is done by using the emitter signals of the ANTARES
acoustic positioning system (see
Subsection~\ref{sec:amadeus_part_antares}).
Figure~\ref{fig:pinger_am_hti} shows such a signal as recorded by four
typical sensors. The delays between the signal arrival times are
clearly visible: short delays of less than 1\,ms within each storey
and a long delay of about 10\,ms between the signals arriving in two
different storeys.

The time shown in the figure is given in seconds since the start of
the run and can be converted into UTC
using the data recorded by the clock system 
(see Subsection~\ref{sec:DAQ-and-clock}). As the emission times of the
positioning signals are also recorded in UTC, the time difference
between emission and reception of the signal can be calculated. 
Using the signals from multiple
emitters and their known positions at the anchors of the lines, the
positions of the AMADEUS sensors can be reconstructed.

The position calibration has statistical uncertainties
of a few millimetres for each hydrophone.  
Systematic uncertainties due to the size of
the receiving piezo elements, the knowledge of their relative
positions within the acoustic storey, the knowledge of the speed
of sound in sea water and the position uncertainties of the emitters
are still under study.  For the AMs, the position reconstruction is
less precise and has a statistical uncertainty of the order of a centimetre.

\section{Summary and Conclusions}
\label{sec:summary}
The AMADEUS system for the investigation of techniques for acoustic
particle detection in the deep sea has been integrated into the
ANTARES neutrino telescope in the Medi\-ter\-ran\-ean Sea at water
depths between 2050 and 2300\,m.  The system started to take data in
December 2007 and was completed in May
2008.  It comprises 36 acoustic sensors, of which currently
34 are operational, arranged in six acoustic clusters. 
Different configurations of sensor clusters are used.
The sensors consist of piezo-electric elements and
two-stage preamplifiers with combined sensitivities
around $-$145\,dB\,re\,1V/$\upmu$Pa.

For the off-shore data acquisition, a dedicated electronics board has
been designed.  One of 12 steps of analogue amplification between 1 to
562 can be set with the on-shore control software.  Data sampling is
done at 500 kSps with 16 bits and an analogue anti-alias filter with a
3\,dB point at a frequency of 128\,kHz.  Digital downsampling with
factors of 2 and 4 is implemented inside an off-shore FPGA. This value
is also selectable using on-shore control software.  The system
parameters were tuned using the data collected with the autonomous
precursor device AMADEUS-0 that was deployed and subsequently
recovered at the ANTARES site in March 2005.

Where appropriate, the components of the AMADEUS system have been
calibrated in the laboratory prior to deployment; the in-situ
performance is in full accordance with the expectations. Data taking
is going on continuously and the data are recorded if one of three
adjustable trigger conditions is met.

The system is well suited to conclude on the feasibility of a future
large-scale acoustic neutrino telescope in the deep sea.  Furthermore,
it has the potential of a multi-purpose device, combining its design
goal to investigate acoustic neutrino detection techniques with the
potential to perform marine science studies.  AMADEUS is a promising
starting point for instrumenting the future neutrino telescope project
KM3NeT~\cite{bib:km3net1,bib:km3net2} with acoustic sensors for
position calibration and science purposes.

\section{Acknowledgements}

The authors acknowledge the financial support of the funding agencies:
Centre National de la Recherche Scientifique (CNRS), Commissariat \`a
l'\'energie atomique et aux \'energies alternatives (CEA), Agence National
de la Recherche (ANR), Commission Europ\'eenne (FEDER fund and Marie
Curie Program), R\'egion Alsace (contrat CPER), R\'egion
Provence-Alpes-C\^ote d'Azur, D\'epartement du Var and Ville de La
Seyne-sur-Mer, France; Bundesministerium f\"ur Bildung und Forschung
(BMBF), Germany; Istituto Nazionale di Fisica Nucleare (INFN), Italy;
Stichting voor Fundamenteel Onderzoek der Materie (FOM), Nederlandse
organisatie voor Wetenschappelijk Onderzoek (NWO), the Netherlands;
Council of the President of the Russian Federation for young
scientists and leading scientific schools supporting grants, Russia;
National Authority for Scientific Research (ANCS), Romania; Ministerio
de Ciencia e Innovacion (MICINN), Prometeo of Generalitat Valenciana
(GVA) and MULTIDARK, Spain. We also acknowledge the technical support
of Ifremer, AIM and Foselev Marine for the sea operation and the
CC-IN2P3 for the computing facilities.

%Bibliography

\bibliographystyle{elsarticle-num}
\bibliography{amadeus_NIM_lahmann}

\begin{thebibliography}{10}
\expandafter\ifx\csname url\endcsname\relax
  \def\url#1{\texttt{#1}}\fi
\expandafter\ifx\csname urlprefix\endcsname\relax\def\urlprefix{URL }\fi
\expandafter\ifx\csname href\endcsname\relax
  \def\href#1#2{#2} \def\path#1{#1}\fi

\bibitem{bib:Askariyan2}
{G.A.\ Askariyan, B.A.\ Dolgoshein \etal}, 
%Acoustic detection of high energy particle showers in water, 
Nucl.\ Instr.\ and Meth. 164 (1979) 267.

\bibitem{bib:Learned}
J.~Learned, 
%Acoustic radiation by charged atomic particles in liquids: An analysis, 
Phys. Rev. D 19 (1979) 3293.

\bibitem{bib:Sim_Acorne}
S.~Bevan \etal\ (ACoRNE~Coll.), 
%Simulation of ultra high energy neutrino interactions in ice and water, 
Astropart.\ Phys. 28~(3) (2007) 366, arXiv:astro-ph/0704.1025v1.

\bibitem{bib:Sim_Acorne2}
S.~Bevan \etal\ (ACoRNE~Coll.), 
%Study of the acoustic signature of UHE neutrino interactions in water and ice,
Nucl.\ Instr.\ and Meth. A 607 (2009) 389,
  arXiv:0903.0949v2 [astro-ph.IM].

\bibitem{bib:Bertin_Niess}
{V.\ Niess and V.\ Bertin}, 
%Underwater acoustic detection of ultra high energy neutrinos, 
Astropart.\ Phys. 26 (2006) 243, arXiv:astro-ph/0511617v3.

\bibitem{bib:spats}
{F.\ Descamps for the IceCube Coll.}, 
%Acoustic detection of high energy
%neutrinos in ice: Status and results from the south pole acoustic test setup,
  in: Proceedings of the 31st International Cosmic Ray Conference, 2009,
  arXiv:0908.3251v2 [astro-ph.IM].

\bibitem{bib:baikal}
{K.~Antipin \etal\ (BAIKAL Coll.)}, 
%A prototype device for acoustic neutrino detection in lake Baikal, 
in: Proceedings of the 30th International Cosmic
  Ray Conference, 2007, arXiv:0710.3113 [astro-ph].

\bibitem{bib:saund}
{J.\ Vandenbroucke, G.\ Gratta and N.\ Lehtinen}, 
%Experimental study of acoustic ultra-high-energy neutrino detection, 
Astrophys.\ J. 621 (2005) 301,
  arXiv:astro-ph/0406105.

\bibitem{bib:acorne}
{S.~Danaher for the ACoRNE Coll.}, 
%First data from acorne and signal processing techniques, 
in: Proceedings of ARENA 2006, the 2nd International
  Workshop on Acoustic and Radio EeV Neutrino detection Activities, 
  J.\ Phys.\ Conf.\ Ser.~81,
  IOP Publishing, Philadelphia, 2007, p. 012011.

\bibitem{bib:noise_ONDE}
{G.\ Riccobene for the NEMO Coll.}, 
%Long-term measurements of acoustic noise in very deep sea, 
in: Proceedings of ARENA 2008, the 3rd International Workshop
  on Acoustic and Radio EeV Neutrino Detection Activities, 
  Nucl.\ Instr.\ and Meth.~A 604, 2009, p. 149.

\bibitem{bib:ANTARES-paper}
{The ANTARES Collaboration}, 
ANTARES, the first operational Neutrino Telescope in the Mediterranean
Sea, to be submitted to Nucl.\ Instr.\ and Meth.\ A.

\bibitem{bib:ANTARES-line1}
{M.\ Ageron et al.\ (ANTARES Coll.)}, 
%Performance of the first antares detector line, 
Astropart.\ Phys. 31 (2009) 277, arXiv: 0812.2095 v1 [astro-ph].

\bibitem{bib:OMs}
{P.~Amram \etal\ (ANTARES Coll.)}, 
%The antares optical module, 
Nucl.\ Instr.\ and Meth. A 484 (2002) 369.

\bibitem{bib:antares-pos}
{M.\ Ardid for the ANTARES Coll.}, 
%Positioning system of the antares neutrino telescope, 
in: Proceedings of VLVnT 2008, the 3rd International Workshop on a
  Very Large Volume Neutrino Telescope for the Mediterranean Sea, 
Nucl.\ Instr.\ and Meth.\ A 602,  2009, p. 174.

\bibitem{bib:Richardt_reco}
C.~Richardt. \etal, 
%Reconstruction methods for acoustic particle detection in the deep
%sea using clusters of hydrophones, 
Astropart.\ Phys. 31 (2009) 19,
  arXiv:0906.1718v1 [astro-ph.IM].

\bibitem{bib:line0}
{M.~Ageron \etal\ (ANTARES Coll.)}, 
%Studies of a full-scale mechanical
%  prototype line for the antares neutrino telescope and tests of a prototype
%  instrument for deep-sea acoustic measurements, 
Nucl.\ Instr.\ and Meth. A 581 (2007) 695.

\bibitem{bib:Deffner_diplom}
{F.~Deffner}, {Studie zur akustischen Neutrinodetektion: Analyse und Filterung
  akustischer Daten aus der Tiefsee},
\newblock  Diploma thesis, Univ.\ Erlangen-N{\"u}rnberg,  2007,
  FAU-PI1-DIPL-07-001, \\ obtainable from
  http://www.antares.physik.uni-erlangen.de/publications.


\bibitem{bib:hoessl2006}
{G.~Anton \etal}, 
%Study of piezo based sensors for acoustic particle detection,
  Astropart.\ Phys. 26 (2006) 301.

\bibitem{bib:Graf_PhD_2008}
{K.~Graf}, 
Experimental Studies within ANTARES towards Acoustic Detection
  of Ultra-High Energy Neutrinos in the Deep Sea,
  Ph.D. thesis, Univ.\ Erlangen-N{\"u}rnberg, 2008,
  FAU-PI1-DISS-08-001, \\ obtainable from
  http://www.antares.physik.uni-erlangen.de/publications.


\bibitem{urick2}
{R.J.\ Urick}, Ambient Noise in the Sea, Peninsula publishing, Los Altos, USA,
  1986, {ISBN 0-932146-13-9}.

\bibitem{bib:naumann_phd}
{C.L.~Naumann}, 
Development of Sensors for the Acoustic Detection of Ultra High
  Energy Neutrinos in the Deep Sea,
Ph.D. thesis, Univ.\ Erlangen-N{\"u}rnberg,  2007,
  FAU-PI4-DISS-07-002, \\ obtainable from
  http://www.antares.physik.uni-erlangen.de/publications.


\bibitem{bib:antares_daq}
{J.A.~Aguilar \etal\ (ANTARES Coll.)}, 
%The data acquisition system for the antares neutrino telescope, 
Nucl.\ Instr.\ and Meth. A 570 (2007) 107.

\bibitem{bib:ARS_paper2}
{J.A.~Aguilar \etal\ (ANTARES Coll.)}, 
%Performances of the front-end electronics of the ANTARES neutrino telescope, 
Nucl.\ Instr.\ and Meth.\ A 622 (2010) 59, 
arXiv:1007.2549v1 [astro-ph.IM].
\bibitem{urick}
{R.J.\ Urick}, Principles of Underwater Sound, Peninsula publishing, Los Altos,
  USA, 1983, {ISBN 0-932146-62-7}.

\bibitem{bib:milom-paper}
{J.A.~Aguilar \etal\ (ANTARES Coll.)}, 
%First results of the instrumentation line for the deep-sea antares neutrino telescope, 
Astropart.\ Phys. 26 (2006) 314.

\bibitem{bib:root}
{F. Rademakers and R. Brun}, Root: An object-oriented data analysis framework,
  Linux Journal 51, http://root.cern.ch/.

\bibitem{Neff_diplom}
M.~Neff, {Studie zur akustischen Teilchendetektion im Rahmen des
  ANTARES-Experiments: Entwicklung und Integration von Datennahmesoftware},
  Diploma thesis, Univ.\ Erlangen-N{\"u}rnberg,  2007,
  FAU-PI1-DIPL-07-003, \\ obtainable from
  http://www.antares.physik.uni-erlangen.de/publications.



\bibitem{bib:richardt_ARENA08}
{C.\ Richardt \etal }, 
%Position reconstruction of acoustic sources with the AMADEUS detector, 
in: Proceedings of ARENA 2008, the 3rd International
  Workshop on Acoustic and Radio EeV Neutrino Detection Activities, 
  Nucl.\ Instr.\ and Meth.\ A 604, 2009, p. 189.

\bibitem{bib:km3net1}
{U.F.~Katz \etal\ (KM3NeT Consortium)}, 
%Status of the km3net project, 
in:
  Proceedings of VLVnT 2008, the 3rd International Workshop on a Very Large
  Volume Neutrino Telescope for the Mediterranean Sea, 
  Nucl.\ Instr.\ and Meth.\ A 602, 2009, p.~40.

\bibitem{bib:km3net2} 
{KM3NeT~Consortium}, Conceptual Design for a
  Deep-Sea Research Infrastructure Incorporating a Very Large Volume
  Neutrino Telescope in the Mediterranean Sea, ISBN 978-90-6488-031-5,
  http://www.km3net.org/CDR/CDR-KM3NeT.pdf (2008).

\end{thebibliography}
%\clearemptydoublepage

\end{document}